\documentclass[astrosymb]{aastex701}

\usepackage{graphics,graphicx}
\usepackage{amsmath}
\usepackage{amssymb}
\usepackage{enumitem}
\usepackage{wrapfig}
\usepackage{xcolor}
\newcommand{\kms}{$\rm km\,s^{-1}$}

\newcommand{\msun}{$\rm M_\odot$}
\newcommand{\lsun}{$\rm L_\odot$}
\newcommand{\lmir}{$\rm L_{MIR}$}

\newcommand{\mir}{MIR}        \newcommand{\mic}{$\mu$m}

\newcommand{\mrs}{{\small MIRI/MRS}}
\newcommand{\htwo}{$\rm H_2$}

\newcommand{\fc}{$f_C$}

\newcommand{\gal}{{NGC$\,$5253}}
\newcommand{\nlyc}{$Q_0$}
\newcommand{\logq}{$\rm log(Q_0)$}

\newcommand{\lmirest}{$\rm L_{MIR}^{est}$}
\newcommand{\tausil}{$\tau_{9.7}$}
\newcommand{\hii}{{H$\,${\small II}}}

\newcommand{\halpha}{H$\,\alpha$}
\newcommand{\paa}{Pa$\,\alpha$}

        \newcommand{\super}{supernebula}        
\long\def\symbolfootnote[#1]#2{\begingroup \def\thefootnote{\fnsymbol{footnote}}\footnote[#1]{#2}\endgroup} 
\graphicspath{{N5253_C8_Images/}}

\usepackage{graphicx}
\usepackage{dcolumn}

\defcitealias{beck2025}{Paper~II}

\begin{document}

\title{JWST View of the Supernebula in NGC 5253. I. Overview and Continuum Features}

\author[0000-0003-4625-2951,gname=Jean L.,sname=Turner]{Jean L. Turner}
\affiliation{UCLA Department of Physics and Astronomy, Los Angeles, CA 90095 USA}
\email[show]{turner@astro.ucla.edu}

\author[0000-0002-5770-8494,gname=Sara,sname=Beck]{Sara C. Beck}
\affiliation{School of Physics and Astronomy, Tel Aviv University, Tel Aviv, Israel}
\email{becksarac@gmail.com}

\author[0000-0001-5678-7509,gname=Elm,sname=Zweig]{Elm Zweig}
\affiliation{UCLA Department of Physics and Astronomy, Los Angeles, CA 90095 USA}
\email{elmazweig@physics.ucla.edu}

\author[0000-0003-0057-8892]{L. Barcos-Mu\~noz}
\affiliation{National Radio Astronomy Observatory, 520 Edgemont Road, Charlottesville, VA 22903, USA}
\affiliation{Department of Astronomy, University of Virginia, 530 McCormick Road, Charlottesville, VA 22903, USA}
\email{lbarcos@nrao.edu}

\author[0000-0001-7221-7207,gname=John,sname=Black]{John H. Black}
\affiliation{Department of Physics and Astronomy, Chalmers Institute of Technology, Gothenberg, Sweden}
\email{john.black@chalmers.se}

\author[]{Daniel P. Cohen}
\affiliation{UCLA Department of Physics and Astronomy, Los Angeles, CA 90095 USA}
\email{daniel.parke.cohen@gmail.com}

\author[0000-0002-0214-0491,gname=Michelle,sname=Consiglio]{S. Michelle Consiglio}
\affiliation{UCLA Department of Physics and Astronomy, Los Angeles, CA 90095 USA}
\email{smconsiglio@ucla.edu}

\author[0000-0002-9064-4592]{Nicholas G. Ferraro}
\affiliation{UCLA Department of Physics and Astronomy, Los Angeles, CA USA}
\email{nferraro@g.ucla.edu}

\author[0000-0002-3412-4306,gname=Paul,sname=Ho]{Paul T. P. Ho}
\affiliation{Academia Sinica Astronomy and Astrophysics, Taipei, Taiwan}
\email{pho@asiaa.sinica.edu.tw}

\author[0000-0001-9436-9471,gname=Davie,sname=Meier]{David S. Meier}
\affiliation{New Mexico Institute of Mining and Technology, Socorro, NM USA}
\email{David.Meier@nmt.edu}

\correspondingauthor{Jean Turner}
\email{turner@astro.ucla.edu}

\begin{abstract}\label{sec:abstract}
We present 5-28\micron\ imaging spectroscopy of the ``supernebula" in the dwarf galaxy 
NGC~5253 made with the JWST MIRI-MRS integral field spectrometer. 
\gal\ hosts a luminous ($L\sim 10^9~\rm L_\odot$) \hii\ region, 
powered by a giant young star cluster,  a possible local analogue to super star 
cluster formation at Cosmic Dawn and Noon. In this paper, the first in a series 
about the mid-infrared line and continuum emission in the center of \gal, 
we present an overview and continuum spectra. The mid-infrared images reveal four 
continuum sources that we identify as  hot dust emission 
from luminous \hii\  regions. 
The dominant source is the  pc-scale supernebula core seen at radio wavelengths.
We find that the MIR to radio continuum flux 
ratio is identical to that of Galactic \hii\ regions. The 9.7\micron\ silicate 
feature is present and strongest in absorption toward the supernebula.  
PAH features at 6.2, 6.8, 7.7, 8.6, and 11.2\micron\ are detected across the region, although  weakly, particularly in the supernebula. 
Unusual PAH line profiles are seen, 
including a broad 6\micron\ feature, likely in part aliphatic, from
the extended starburst region and strong 11.0\micron\ emission from the supernebula. These observations suggest dust processing 
by intense radiation fields and possible dust production and 
dispersal by Wolf-Rayet stars within the region.
\end{abstract}

\section{Introduction} \label{sec:intro}

Giant star clusters may have been the dominant mode of star formation in the early universe, 
responsible for dusty galaxies and rapid heavy element enrichment,  notably of nitrogen,
at early times  \citep{cameron2023, klessenglover2023, dayalferrara2018}. Massive enough to contain thousands of O stars, giant star clusters are critical sources of radiation, luminosity, and feedback activity that serve to shape the host galaxy. Despite containing large numbers of active high-mass stars, these clusters not only lived to great ages to become globular clusters, but also continued to form multiple generations  \citep{bastianlardo2018}. What determines the limit to the size of a star cluster? How do they continue to form stars in the presence of winds and supernovae? How and when do they enrich their surroundings? 
Clusters as massive as $\sim 10^6$~\msun\ are not currently forming within the Local Group; Arches Cluster and even 30 Doradus are more than an order of magnitude smaller. To find potential protoglobular clusters in formation requires extragalactic observations. The James Webb Space Telescope (JWST), with subarcsecond capability in the mid-IR, is ideal for observing the formation of these rare regions.

An excellent candidate for the study of the formation of a giant 
cluster in a low metallicity environment  showing evidence of nitrogen enrichment  
is the  \hii\ region in the nearby (3.7 Mpc) dwarf galaxy, \gal. 
By kinematics and overall morphology, \gal\ is classified as a peculiar dwarf spheroidal  \citep{caldwell1989}. 
It has metallicity $Z\sim 0.3~\rm Z_\odot$ \citep{walshroy1989,kobulnicky1997,monreal-ibero2010}, similar to the Magellanic clouds \citep{dufour1984, russell1992}.  
The similarity of the giant nebula in \gal\ to 30 Doradus was noted 
in the first spectrum ever obtained of the source \citep{pickering1896}.
The galaxy contains dozens of massive star clusters 
\citep{caldwell1989,gorjian1996,calzetti1997,harris2004,cresci2005,
degrijs2013,calzetti2015,smith2016}, 
the brightest of which are blue and young, with ages of 1-30 Myrs 
\citep{calzetti1997,harris2004,degrijs2013}. 
Extremely bright [S~IV] emission \citep{aitken1982,moorwoodglass1982,beck1996} 
  indicates that a giant cluster containing very massive stars must be present \citep{aitken1982,moorwoodglass1982,beck1996,crowther1999}. Nitrogen
  enhancement near the [S~IV] source and related Wolf-Rayet spectral features have long  been noted 
  \citep{walshroy1989,kobulnicky1997,schaerer1997,monreal-ibero2010,pruijt2026}. 

High resolution imaging with the Very Large Array (VLA) \citep{beck1996} revealed
a compact ($\sim 1$-3 pc) radio source, the ``supernebula",
corresponding to an \hii\ region far  
more luminous than indicated by \halpha\ emission \citep{calzetti1997}.  
The radio and infrared data suggest a Lyman continuum rate for the pc-scale supernebula core consistent
 with a cluster of $\gtrsim$1000 O stars and $L\sim 10^9~\rm L_\odot$, 
 \citep{beck1996, turner1998,rodriguez-rico2007}. 
Discrepancies between \halpha\ and radio and infrared line fluxes 
indicate high visual extinctions, $A_V \sim 33$-35~mag  
\citep{calzetti1997,kawara1989, turner2003,alonso-herrero2004, martinhernandez2005} toward this cluster. 
The region of nitrogen enhancement is located within a few pc to the north of the supernebula \citep{pruijt2026}.

Despite its population of young stars and extensive recent 
star formation \citep{harris2004}, \gal\ is unusual in that there is
 little molecular gas within the galaxy proper, particularly near the supernebula 
\citep{turner1997,meier2002}.  
The discrepancy is unlikely to be explained by CO-dark gas, which dominates
on larger scales \citep{hunt2023} but typically not in denser, star-forming regions.
CO imaging suggests that the star formation is being
fueled by stream accretion from a redshifted streamer outside the
galaxy along the prominent dust lane 
\citep{turner1997,meier2002,turner2015,turner2017,consiglio2017, miura2018}.
\gal\ is surrounded by a halo of filamentary neutral hydrogen 
\citep{kobulnicky1995,kobulnicky2008,lopez-sanchez2012}. The ``ionization
cone" first noted by \citet{graham1981} 
appears to be on the surface of this infalling streamer; its excitation
probably arises from the central starburst \citep{zastrow2011}, 
and it is more likely an ``ionization sheath" around the molecular gas.

 Based on previous mid-infrared spectra revealing high excitation nebular lines
\citep{verma2003,martinhernandez2005,beirao2006},
there is speculation that very massive stars (VMS) could be present in the
giant \hii\ region in \gal,
with masses perhaps as high as 300 \msun\ \citep{crowther1999,smith2016}. 
It is also possible that, in
such a massive cluster,  self-enrichment could
occur so quickly that catastrophic cooling could stall stellar
winds and perhaps even supernova remnant expansion 
\citep{silich2004,tenoriotagle2005}, which could limit feedback. The evolution of
the giant \hii\ region in \gal\ is, for these reasons, a target of
special interest.  Because of the apparent high visual extinction, 
the need for a subarcsecond resolution
mid-infrared (\mir) picture of this cluster using JWST is clear.

This paper is the first of a series describing
imaging spectroscopy of the giant star cluster and nebula in \gal\ with 
the Medium Resolution Spectrometer of the  Mid-Infrared Instrument (MIRI) 
\citep{wells2015,argyriou2023a}. These observations
give a glimpse into the innermost workings of a giant star cluster 
and its environment while it is in the process of active formation.
In Paper I, we provide an overview of the observations and 
data reduction  in $\S$\ref{sec:obs}. 
In $\S$\ref{sec:results} 
we define the spatial regions that will later be used as
apertures for line flux determinations, and discuss
the continuum spectra and mid-infrared luminosity 
contributions. We discuss the 9.7\micron\ silicate feature in 
$\S${\ref{subsec:res_silicate}} and 
PAH features in $\S$\ref{subsec:res_PAHs}. 
In Paper II \citep{beck2025} we discuss the nebular and molecular
lines for the regions defined here.

\section{Observations and Data Reduction} \label{sec:obs}

The Cycle 1 program (JWST GO-2402, P.I. J. Turner) was executed in 2022 March. It  
consists of a single science pointing of the MIRI/MRS spectrometer and a 
dedicated background observation. 
The design and flight performance of MIRI/MRS are 
described in \citet{wells2015} and \citet{argyriou2023a}. 
Data were taken in each of the four 
channels and at each of the three grating settings 
(short, medium, and long) for a total of twelve bands spanning the wavelength
range 4.9-28\mic.  Observations were executed using a 4-point 
dither and FASTR1 readout mode, with 4 integrations of 16 seconds and a total exposure
time of 477 seconds. The dedicated background was observed for an equal amount
of time.  The spatial resolution of MIRI/MRS 
ranges from 0.4\arcsec\ ($\sim 7$~pc) at 5\mic\ to 1.0\arcsec\ ($\sim 18$~pc) 
at 27\mic, full width 
half maximum (FWHM) for drizzled cubes; spectral resolution ranges from 1300-3700, with some variation within bands \citep{argyriou2023a}. 

\begin{figure}
\begin{center}
\includegraphics[width=3in]{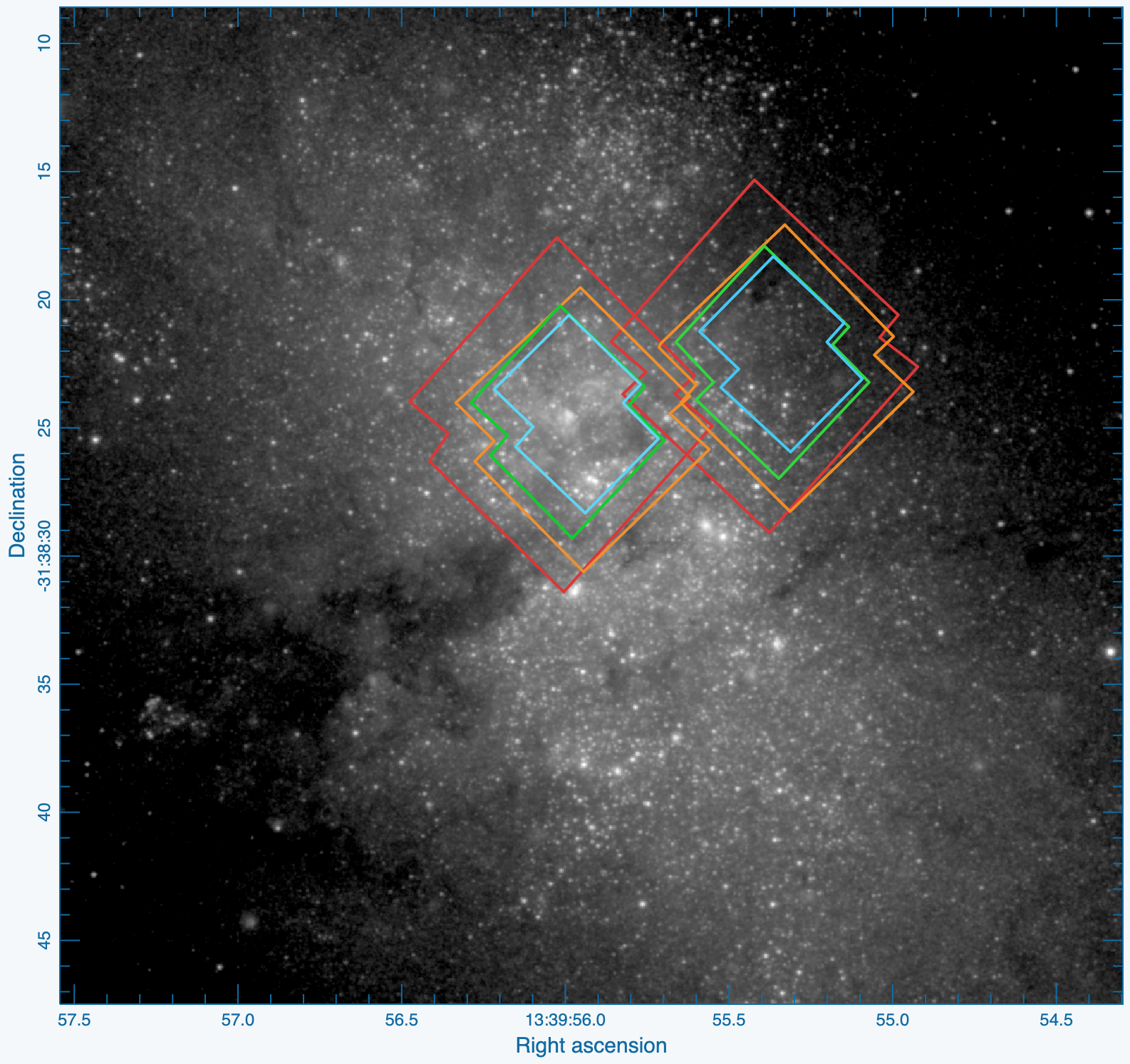}
\end{center}
\caption{Footprints for the JWST \mrs\ cubes
for the science (left) and background (right) observations.  North is up, east is left. }
The science footprint is centered on the bright radio ``supernebula" near
the brightest visual source. 
Footprints are shown for the four channels: Bands 1-S (4.95\mic; blue), 2-L 
(near [SIV]10.51\mic; green),
3-L (near [Ne III]15.55\mic; orange) and 4-L (26.0\mic; red). The 
southeastern corner of the background has continuum and weak line 
emission beyond 12\mic\ in Channels 3 and 4. 
\label{fig:footprints}
\end{figure}

The science pointing was centered on the radio core in \gal, which is located at
R.A. 13$^h$39$^m$55\fs 965, Dec. {-31\degr} 38\arcmin 24\farcs 364 (ICRS). In Figure \ref{fig:footprints} we show the detector
footprints of the four channels for both the science observations
and the background on an HST/ACS F555W image of \gal. 
The footprints shown span the four channels, from Band
1-S at 4.9\mic, to the longest wavelength, Band 4L at 28\mic.
Science and background footprints overlap 
in Channel 4. 

NGC 5253 is uncommonly bright in the infrared, so bright that
multiple reflections within the detector
gave rise to strong fringing and spatial artifacts. In addition
to the ``petals" of the point spread function (PSF), there are artifacts
such as the cruciform \citep{gaspar2021} that 
affect the entire footprint.  

Data were processed by the standard pipeline \citep{gordon2015}  version
1.19.1.  Pipeline runs were performed on the science and background data 
separately with the MRS\_FlightNB1 Jupyter notebook \citep{law2025}. 
Default pipeline parameters were used, with no background subtraction performed
in the pipeline.

\begin{figure}[h]
\begin{center}
\includegraphics[width=\textwidth]{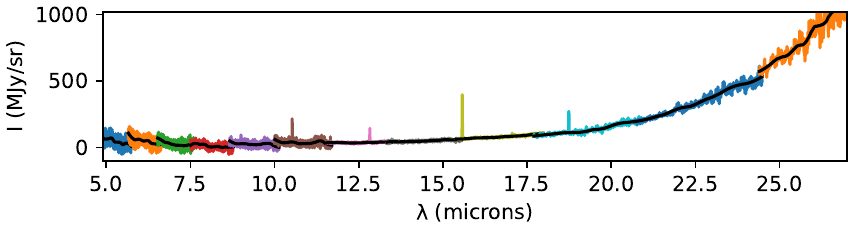}
\end{center}
\caption{
 Custom background spectrum, computed from clean regions within the dedicated background.
Mean intensity levels correspond to fluxes  of $\sim 50$mJy for $\lambda < 20$\mic,
rising to $\sim$ 0.5 Jy at 25\mic, for the region corresponding
to the footprint at 5\mic. For the 0\farcs8 apertures used for spectral extraction, 
background fluxes range from 2-10 mJy from 5-28\mic. 
Spectral lines were removed by boxcar averaging and the spectrum was loess-smoothed
(black line) prior to subtraction from the science cube.}
\label{fig:backgrounds}
\end{figure}

The dedicated background contained emission within the southeast portion of
its footprint. We created a custom background using spectra from 
smaller regions relatively free of emission within the background footprint. 
Since the same integration time was devoted to background as to the science target,  there was adequate sensitivity. We selected elliptical regions 
to the northwest edges of the footprint of extent $\sim$2\farcs5 x 1\arcsec\ 
(FWHM), with shifts in the different bands to accommodate
the varying band footprints. 
 A 1D background intensity spectrum was constructed from the custom region 
for each band using the ``MEAN" statistic of CARTA  \citep[][]{CARTA}; the
spectrum is shown in Figure \ref{fig:backgrounds}. 
The background spectrum was loess-smoothed in R as described below. 

Residuals between the loess-smoothed fit and the background spectrum
are less than 0.6-0.7\% and include contributions by weak lines as well
as steps in continuum levels among the 12 bands. 
Following the ``master background" method of the pipeline, 
the resulting smoothed line-free 1D custom background 
spectrum was broadcast over the 2D spatial footprint of the science data 
to form a background cube, uniform across the spatial dimensions,
varying along the spectral axis. 
Intensities were significantly reduced in the custom background by up to 75\% over 
the default master background, to mid-band levels of 
20--25 MJy/sr for $\lambda<13$\mic, $\sim 40$-70 MJy/sr for $13$-16\mic, 
$\sim$100-200 MJy/sr for $\lambda \sim 17$-21\mic, and 
$\sim$400-660 MJy/sr  for $\lambda\sim 22$-25\mic, 
 beyond which the response degrades significantly. The background is
slightly lower in Band 1L, 5-10 MJy/sr at 7-8\mic.
These intensities correspond to roughly 10\% of the continuum intensity for 
fainter sources D2 and D6 in these bands, and $\lesssim$ 2\% of
the D1 continuum in Channels 3 and 4.  In terms of flux,  
the background contributes $\lesssim$0.05-0.1~Jy in Channels 1 and 2, 
and $\lesssim$0.4-0.6 Jy in Channels 3 and 4, respectively, 
over the entire 41.6 square arcsecond detector footprint ``Chip" region.
The background flux is $\sim$20 times less 
than these values in the smaller 
0\farcs8 spectral extraction regions discussed below. As shown in
Figure~\ref{fig:backgrounds}, the background spectra of
the shorter wavelength bands in Channels
1 and 2 appear to be dominated by instrumental response. 

ICRS coordinates were assigned to the cubes by comparison with a high resolution 
1.3mm continuum image from the ALMA archive (P.I. D. Nguyen), 
which is largely free-free emission  \citep{turner2017,consiglio2017}. 
Shifts were based on the two strongest mid-IR continuum sources using
a 5.0\mic\ continuum image constructed from the cube.  Shifts applied to the cubes were
$\sim$ -0\farcs3 in R.A. and +\farcs04 in Dec. The
astrometry is estimated to be good to $\lesssim 50$mas. 

The emission is a combination of point and extended sources, 
so rather than using the ``cone" method of flux extraction we
defined fixed spatial apertures for 
studying spectral variations across the footprint. 
For \mrs\ it is recommended \citep{law2023} that an aperture larger
than the observed point spread function is used 
to compensate for aperture broadening, undersampling,  bigger-fatter,
and  pull-up/push-down 
effects \citep{argyriou2023b,law2023,dicken2024}. These
artifacts are all evident in the science data in \gal. Many
of these artifacts are due to multiple reflections within the detector,
which are significant for this very bright source. 
As a result, it is 
not possible to do pixel-by-pixel spectra from the cubes. 
Instead, for the flux measurements we chose 
spatial regions of interest and defined apertures 
of radius $\sim$0\farcs9, 
the FWHM expected for the longest wavelengths.
 For the PAH analysis  slightly smaller apertures, 
 were used, 0\farcs6-0\farcs8. 
Formal aperture correction factors are $\lesssim$15\% \citep{argyriou2023a}. 

For each spatial region (discussed in $\S$\ref{sec:results}), 
we obtained 1D spectra for the twelve band cubes
with CARTA, using the ``SUM" statistic. 
The 1D band spectra were corrected for fringing \citep{gasman2023} using 
the standalone package fit\_residual\_fringes\_1d
available from  jwst.residual\_fringe.utils.
\footnote{
\url{https://jwst-docs.stsci.edu/known-issues-with-jwst-data/miri-known-issues/miri-mrs-known-issues\#gsc.tab=0}}

There were minor spectral mismatches of $\lesssim 1$\% between adjacent overlapping bands for most of the spectrum, with 
larger values of $\sim 1$-7\% for Band 4 ($\lambda > 18$\micron). 
Various procedures have been devised to produce
a smooth spectrum \citep[e.g.][]{chown2023, lai2024, canin2026}.
Since the offsets were small and the spectra relatively smooth (weak
PAH features) we used the following method. 
We first determined the original pipeline 
luminosity of each region by integrating the spectrum before stitching. 
The overlap regions in the spectra were spliced at the midpoints, with offsets
 added to match the mean line-free continuum values in the
overlap regions. After the levels
were matched, a constant offset was added to all bands to restore
the original pipeline-calibrated luminosity for Channels 1-3. Channel 4 was then matched without luminosity adjustment; the final
luminosity differs by $\lesssim 2.5$\% from the original
pipeline luminosity for all spectra except the D4 region, as noted below.

To fit  continuum and broad solid state features in a spectrum marked 
by strong and narrow nebular lines, we performed a non-parametric weighted 
local regression (loess) smoothing of the 1D spectra using numpy fit-loess. 
Weights were chosen to omit all lines from the fits; this
was feasible since the nebular lines are essentially unresolved
\citepalias[identification and analysis is discussed in][]{beck2025}.
The loess smoothing consisted of the default degree two polynomial and 
tricubic weighting on the full band-matched, defringed 1D spectrum for each region, 
using a span of 1\% over the 11625 spectral points. This smoothing is equivalent to reducing resolution to 
$R\sim 200$ at  the shorter wavelengths and $\sim 100$ for $\lambda > 15\micron$. 
The loess-smoothed continuum is an excellent fit to the line-free continuum, with 
residual standard errors  $\lesssim 1$\% for $\lambda <26$ \micron. We do
 not show the loess model for the continuum since it is nearly indistinguishable from the data at print resolution.

Loess-smoothed line-free spectra were also used
for the modeling of the profiles of the broad 9.7\micron\ silicate feature and the PAH features discussed in $\S3.7$. 
Fits of profiles of  individual solid state features
using modified Drude and Gaussian functions are described in detail in the Appendix.

\section{Results} \label{sec:results}

\subsection{The Supernebula Landscape: Gas, Dust, and Clusters}\label{subsec:res_landscape}

\begin{figure}[h]
\begin{center}
\includegraphics[width=4in]{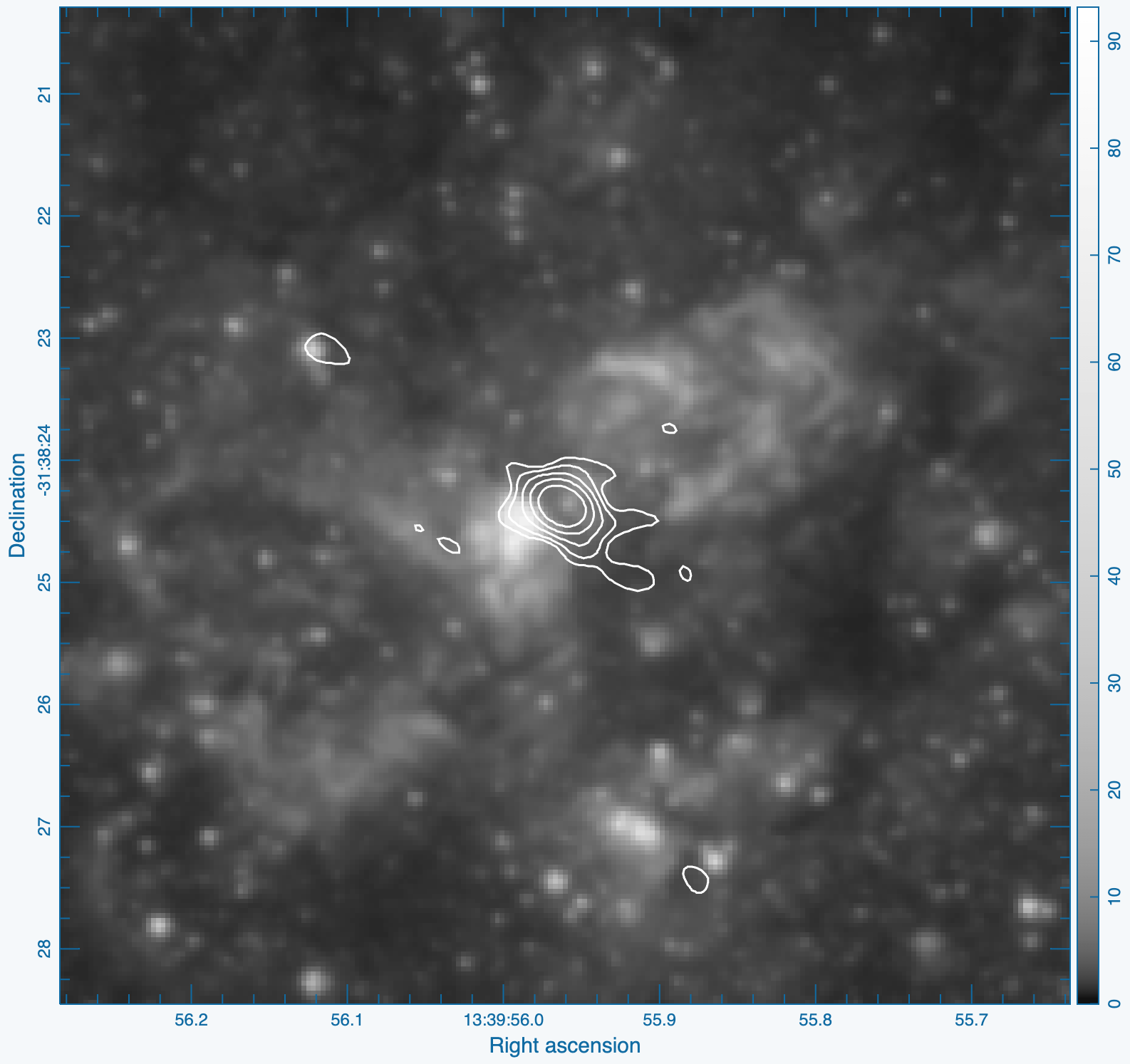}
\end{center}
\caption{
HST-ACS F555M image (Project 10765, P.I. Zezas) with ALMA 1.3mm continuum (Project 
2017.1.00964.S, P.I. Nguyen) 
contours. The 1.3mm continuum is dominated by 
free-free emission at this resolution. Cluster 5 
\citep{calzetti1997} 
is the brightest visual source, located $\sim 0.25$-0.3\arcsec\ to
the southeast of the radio supernebula; Cluster 11 is faint and visible
near the center \citep{smith2020} Registration of the HST image to ICRS coordinates 
was done by aligning nearby Gaia stars \citep{cohen2018}, also \citep{smith2020}. 
Coordinates are shown, north is up, east is to the left.}
\label{fig:clusterfive}
\end{figure}

Like a cubist painting, \gal\ presents different faces to observers 
depending on wavelength. The radio view of \gal\ is very different from the 
optical view. The brightest source at visible wavelengths in \gal\ is 
 ``Cluster 5" of 
\citet{calzetti1997,calzetti1999,calzetti2015} or ``Cluster 1" of \citet{harris2004}.  
Radio imaging reveals a bright radio core, the supernebula, $<$1 pc in 
radius \citep{beck1996, turner1998,turner2000,turnerbeck2004}, 
marking the location of the luminous and compact star cluster.
The radio core, shown in Figure~\ref{fig:clusterfive}, 
lies in a dust lane, close to, but slightly offset (0\farcs2, 3-4 pc) from, a faint optical
cluster, Cluster 11, and $\sim 0.4$\arcsec\ from Cluster 5 \citep{smith2020}. 
Bipolar plumes of emission in \halpha\ \citep{calzetti1997} emerge from the center of the \hii\ region to cover a region more than
100 parsecs in extent, criss-crossed by dust lanes. 
For comparison, the Great Nebula in Orion, excited by a single O star, 
spans about 10 pc, and 30 Doradus nebula in the LMC,  a less luminous \hii\ region,  extends more than 250 pc. 
The total hydrogen ionization rate implied by the radio flux is 
$Q_0 \sim 7\times 10^{52}~\rm s^{-1}$, of which $\sim 30$\% arises
in the radio core 
\citep{turner1998,mohan2001,turnerbeck2004,mohan2001,rodriguez-rico2007,
bendo2017,turner2017}. This value of $Q_0$ is 
consistent with the strong [S~IV] 10.5\mic\ emission 
first seen in early mid-infrared spectra  \citep{aitken1982,moorwoodglass1982,
beck1996}.

The infrared counterpart to the supernebula was 
detected in subarcsecond mid-IR imaging at Keck and its flux confirmed 
that the radio source was likely to be a compact \hii\ region \citep{gorjian2001}.
The IR source, coincident with the radio source \citep{cohen2018, alonso-herrero2004} 
is invisible to wavelengths shorter than
$\lambda\sim$1.9\mic\ \citep{turnerbeck2004,alonso-herrero2004},
consistent with  $A_V\sim 35$ \citep{calzetti1997, beck2025}.
The IR luminosity of the 100 pc region is estimated to be $\sim 10^9~\rm L_\odot$ 
\citep{vanzisauvage2004,hunt2017}.

The radio-MIR  \hii\ region is coincident with a compact ($r<3$~pc)
 CO(3--2) cloud that is unusually warm and chemically peculiar \citep{turner2015, turner2017, consiglio2017, andersonphd2016}. The CO probably arises
in photodissociation regions (PDRs) associated with multiple molecular clumps 
orbiting with the stars within the \hii\ region \citep{turner2015,turner2017}.
 Dust within this CO cloud could explain
the high extinction to the radio core/IR cluster. However, based 
on extended radio emission \citep{rodriguez-rico2007} and the bipolar plumes
evident in H$\alpha$, it is almost certain that the CO cloud is porous \citep{turner2017,consiglio2017}, allowing photon escape.  We discuss 
further JWST \mrs\ evidence for the photon escape in \citetalias{beck2025}.

\citet{beirao2006} presented a \mir\ spectrum
from Spitzer of the central region of \gal. The Spitzer-IRS spectra
indicated high line excitation, increasing toward the central
starburst region \citep{wu2006}. However, the spatial
resolution of Spitzer was unable to resolve individual
pc-scale clusters, and the \mir\ continuum images were strongly 
affected by the central continuum source. Aside from the trend  of excitation appearing to increase from 30 to 300 pc,
the slit was too wide to attribute spectral features specifically
to the \super, although the spatial trend  suggested it was the source
of the high excitation.  The Spitzer results suggested that
higher spatial resolution would allow a more complete understanding
of this complex region.

\subsection{JWST Isolates the HII Regions within a Dusty Starburst}\label{subsec:res_regions}

High resolution HST, Keck, and ALMA observations show multiple optical, radio, and molecular line sources within the inner few arcseconds. The goal of this project was to isolate 
the embedded cluster at the core of the \super\ and observe its effects on
its surroundings at cluster (5 pc) spatial scales.
 Mid-infrared continuum imaging is crucial for observing and
 characterizing the surroundings of the embedded cluster in this dusty region. 
 JWST \mrs\ has the spatial resolution to resolve these structures at the
subarcsecond level. Because \mrs\ is a spectrometer, \mir\ continuum images can be 
constructed for a source that would otherwise be far too bright for JWST.  

\begin{figure}{h}
\begin{center}
\includegraphics[width=\textwidth]{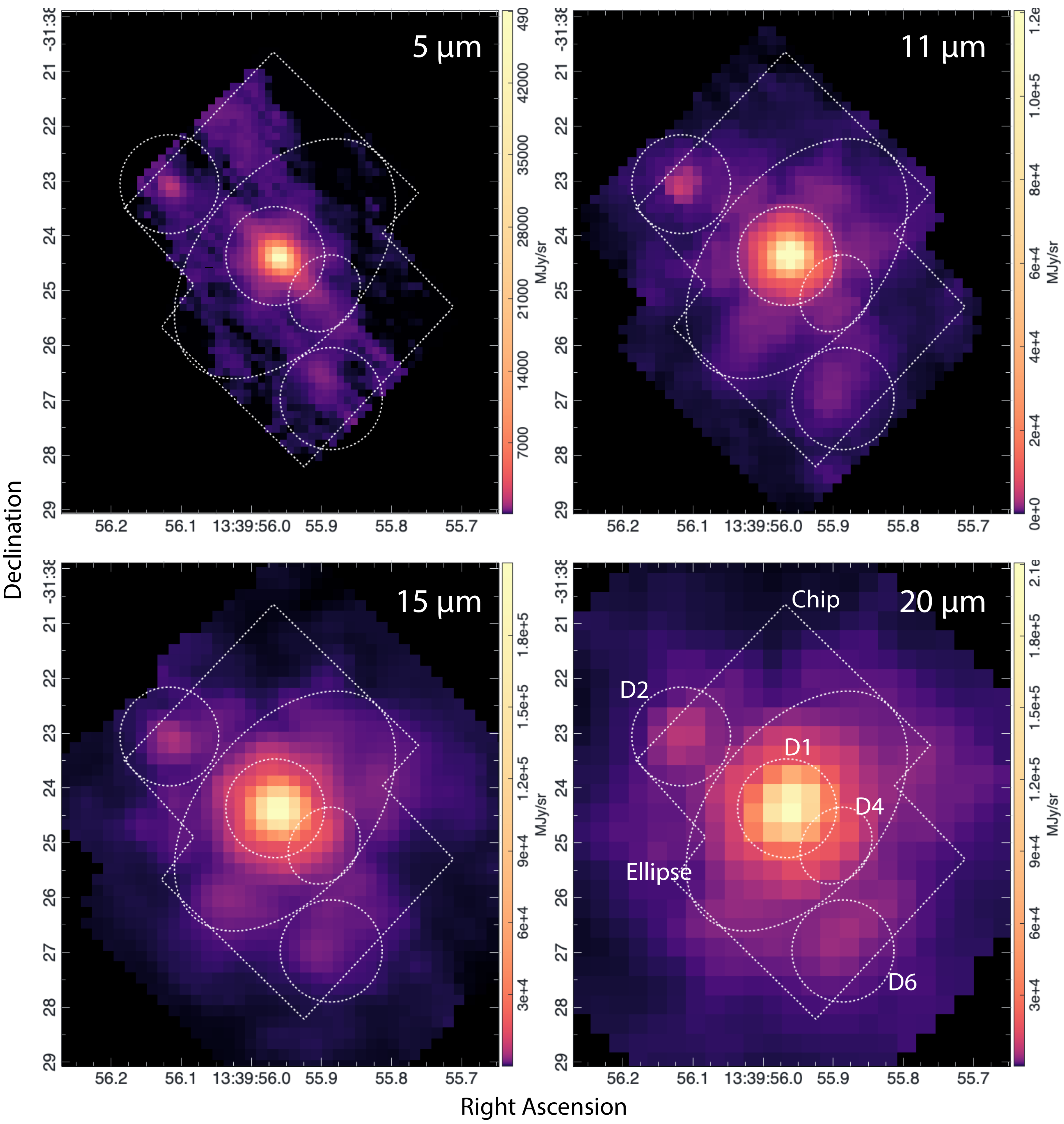}
\end{center}
\caption{MIRI-MRS continuum images, constructed from median per pixel averages 
over the specified wavelength ranges. Regions from which fluxes were extracted
are shown.  The smaller regions have radii of 0\farcs9, 15 pc. 
(Top left): 4.95-5.0\micron. (Top right): 10.6-10.8\micron. 
(Bottom left): 14.7-15.0\micron. (Bottom right): 19.5-20\micron. Plots are to the
same scale, in ICRS coordinates. Regions from which spectra
were extracted are identified in the bottom right panel. }
\label{fig:Continuum4panel}
\end{figure}

\mir\ continuum images  of the central 100~pc starburst region of \gal\  
are presented in Figure \ref{fig:Continuum4panel} for each of the four channels at 
wavelengths ranging from 5\mic\ to 20\mic.  Overlaid are contours 
corresponding to the spectral extraction regions discussed below. 
The continuum images were constructed from median averages of
line-free regions (at R=$\lambda/\Delta \lambda \sim 100$)  from the \mrs\ cubes. 
Even at spectroscopic resolution, JWST is so sensitive that 
the brightness of the \super\ source affects data across
the \mrs\ footprint and imaging artifacts are evident: 
the ``petals" of the PSF around the central \super\  
and the cruciform artifact due to reflections within the detector 
\citep{gaspar2021}  running from northeast to southwest. 
Multiple reflections within the detector also broaden the PSF 
\citep{argyriou2023a,law2023}. 

\begin{figure}[h]
\begin{center}
\includegraphics[width=\textwidth]{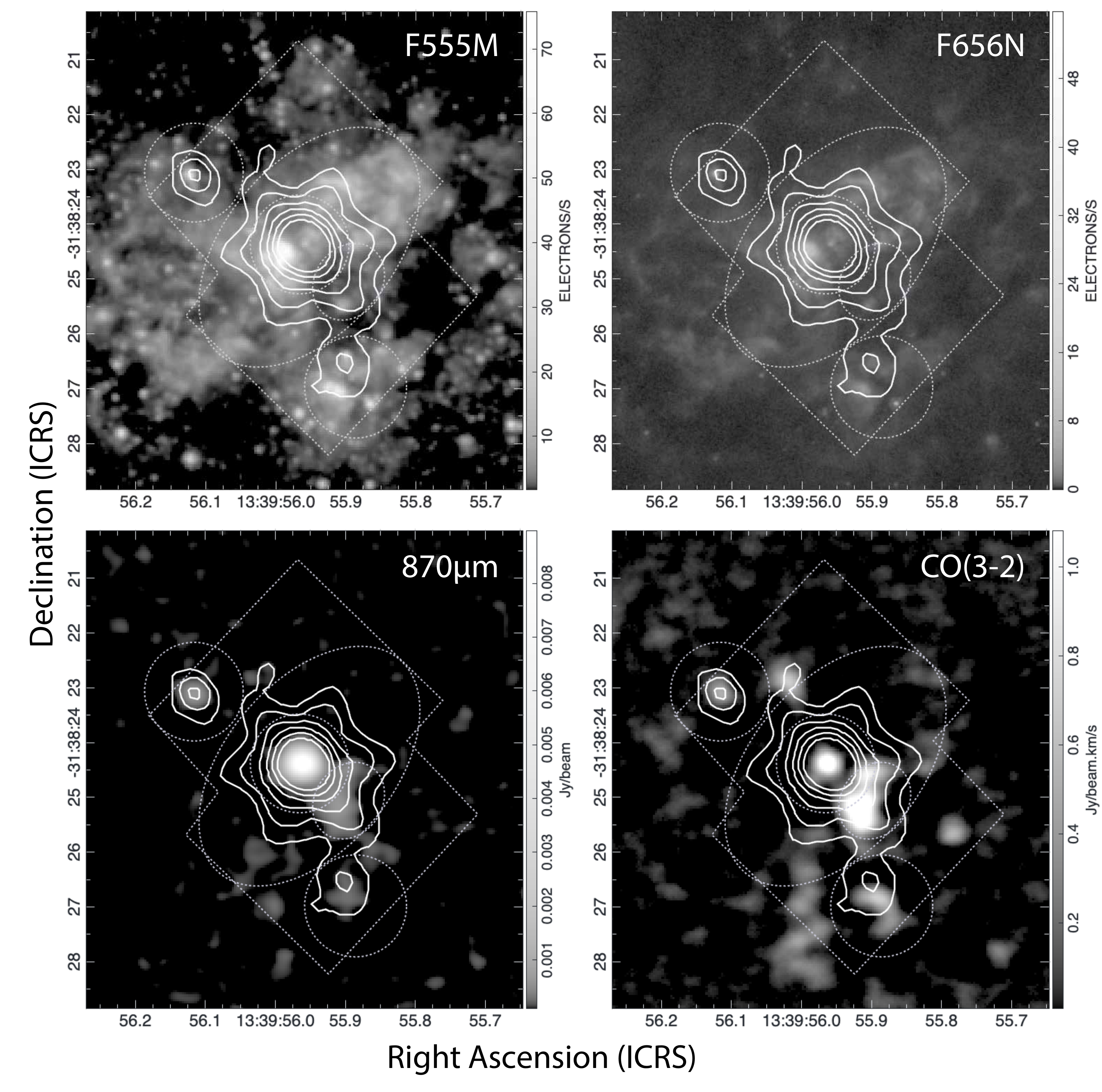}
\end{center}
\caption{
JWST 7\mic\ continuum contours atop images at other wavelengths.
Contour levels of the 7\micron\ image are $240 \times 2^n$ MJy/sr. Astrometry
for the HST images was done using Gaia \citep{cohen2018}, and for the MRS images
from comparison to ALMA 1.3mm images, as discussed in the text.
Peak intensity at 7\mic\ is $1.24 \times 10^5$~MJy/sr. The contours 
are overlaid on:
{\it(Top left):} HST/ACS F555M;  
{\it(Top right):} HST/ACS F656N (``H$\alpha$");
{\it(Bottom left):} ALMA 870$\mu$m continuum;  
{\it(Bottom right):} ALMA CO(3-2) integrated intensity. Coordinates are ICRS, 
with north up, and east to the left. The two
ALMA images \citep{turner2017, consiglio2017} have the same FWHM as MIRI/MRS at 7\mic.}
\label{fig:7micron4panel}
\end{figure}

The \mrs\ continuum  reveals three distinct continuum sources  
in \gal\ within the \mrs\ footprint, and a less distinct fourth. 
These can be seen clearly in Figure~\ref{fig:7micron4panel}, 
which shows images constructed from the median values of the
1L cube at 7.0-7.05\mic\  as contours
overlaid on visible and submillimeter images.

\begin{figure}
\includegraphics[width=\textwidth]{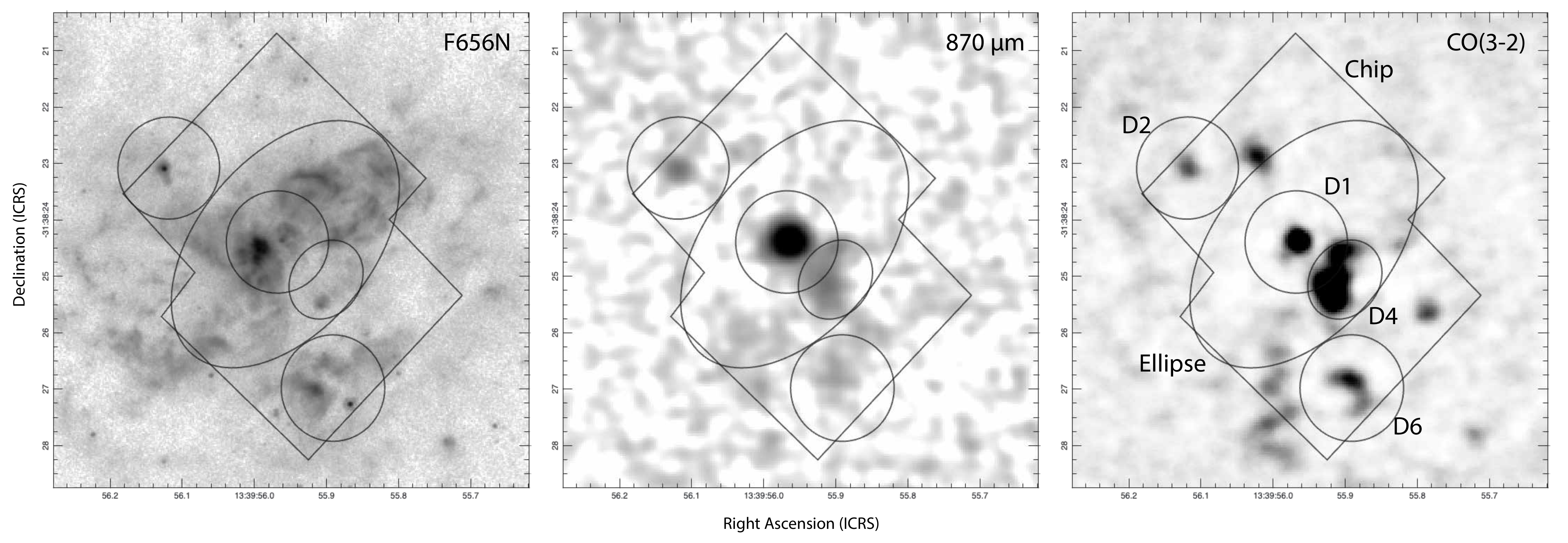}
\caption{
Regions for spectral extraction as viewed at other wavelengths. 
The smaller regions have radii of 0\farcs9, 15 pc; the Chip region is 7\arcsec x
6\arcsec, $\sim 125$~pc $ \times 110$ pc. 
 {\it(Left):} HST F656N; coordinates determined by reference
to GAIA stars \citep{cohen2018}. Region D1 encompasses both the radio 
core and brightest optical source. Ionized plumes extending to northeast
and southwest are captured in the ``Ellipse".
{\it(Middle):} ALMA 870$\mu$m continuum emission. D1 and D2 are dominated by free-free
emission. The D4 region has significant submillimeter dust emission.
{\it(Right):} ALMA CO(3-2) emission with region names based on  
 CO cloud designations of \citet{consiglio2017}.}
\label{fig:regions}
\end{figure}

 Regions from which spectra were extracted are shown in Figures~\ref{fig:Continuum4panel}
and \ref{fig:regions}.  Region names beginning with 
``D" are named for molecular clouds, as identified in Owens Valley and ALMA CO maps
of similar spatial resolution to JWST \citep{meier2002, consiglio2017}.
Although they bear CO names, the spectral regions are not centered on the CO 
centroids  and are generally larger than the CO clouds. 
Coordinates for spectral extraction regions are given, as well as
positions from the high resolution 1.3mm ALMA image of
Figure \ref{fig:clusterfive}. These coordinates are 
listed in Table~\ref{tab:cdetails}. The individual regions 
corresponding to the spectral apertures are the 
following.

1) {\it Region D1, the core of the \super}. Region center: 13$^h$ 39$^m$ 55\fs 966, $-31^\circ$ 38\arcmin\ 24\farcs37, aperture radius: 0\farcs9. Figure~\ref{fig:7micron4panel} 
shows that  the dominant IR continuum  emission arises from the
compact, luminous \super\ core that lies within a dust lane visible
in the HST 814W and in the 656N (H$\alpha$) filters. The
dominant infrared source is coincident wih the radio core
\citep{alonso-herrero2004, cohen2018}. We call this source NGC5253-D1 (``D1"). 
Plumes of emission emerging from this source are visible
 in optical continuum and \halpha. The plumes also appear in the 
 JWST continuum images, although the inner portions unfortunately  
coincide with the ``petals" of the PSF.  
Based on radio free-free and recombination line fluxes,
D1  contains a large star cluster with $\sim$1000 O stars. The CO(3-2) emission in D1 
is unusually hot and apparently optically thin \citep{consiglio2017}. 
These peculiarities in the CO are strong evidence that the
molecular gas in D1 is mixed with the nebula within  $r\lesssim 3$~pc (0\farcs 2). 
\citep{turner2017, consiglio2017}. The CO cloud and
dense core of the \hii\ region are smaller than the spectral region
defined here,  $\sim$0\farcs1-0\farcs2 or $\lesssim2$-3~pc, FWHM \citep{turnerbeck2004,turner2017}.

2) {\it Region D2, pointlike \hii\ region, possible blister}. 
Region center: 13$^h$ 39$^m$ 56\fs117, $-31^\circ$ 38\arcmin\ 23\farcs06; aperture semi-major
axes: 0\farcs9. The CO centroid is 13$^h$ 39$^m$ 56\fs 11, $-31^\circ$ 38\arcmin\ 23\farcs 1 (1.3mm). 
D2 is located 2\farcs 3 ($\sim 40$~pc in projection) to the northeast from D1.
It is bright in the HST/NICMOS \paa\ image of \citet{alonso-herrero2004}.
Although the IR flux of the region corresponding to D2 is 30 times less than the flux of D1, it is
 a luminous \hii\ region in its own right, 
comparable to the luminous Galactic \hii\ regions NGC~3603 or W49A, 
although D2 is smaller, $\lesssim 5$~pc in extent. Since it has a visible H$\alpha$ region
(Fig.~\ref{fig:regions}), D2 may be more evolved than D1, perhaps a blister \hii\ region.
 D2 has also been identified as a region of enhanced He I 6678 emission with
a broad linewidth by
\citet{pruijt2026}, which they identify with cluster G106 of \citet{degrijs2013}. 
This suggests that D2 contains Wolf-Rayet stars.
D2 overlaps the edge of the chip at 5\mic\  slightly but since D2 is a point source, 
its PSF falls mostly within
the chip.

3) {\it Region D4, giant molecular cloud and \hii\ region near D1}. 
Region center: 13$^h$ 39$^m$ 55\fs 896, $-31^\circ$ 38\arcmin\ 25\farcs05, 
aperture radius: 0\farcs74 x 0\farcs61, 325d.
 Region D4 is defined specifically to study the effects of D1 on 
neighboring molecular clouds. D4 is associated with the  most massive CO cloud at 
13$^h$ 39$^m$ 55\fs 92, $-31^\circ$ 38\arcmin\ 25\farcs 0, 
roughly 0\farcs 9 (17 pc in projection) to the west of D1. 
{D4  has an \htwo\ mass $M_{H_2}\sim 10^6$ \msun\ (including He), 
at least 20 times that of the D1 cloud; it is redshifted by 15 \kms\ with respect to 
D1 \citep{rodriguez-rico2007, turner2017, consiglio2017}.
Most of the bright 870\mic\ continuum (Fig.~\ref{fig:regions}) 
in D4 is due to dust, although there is a small
free-free component \citep{consiglio2017} and visible \halpha. }
D4  is less distinct in \mrs\  continuum
images because it is near the petals of the PSF from D1. 
Because it is so close to D1, the D4 region overlaps with D1 at wavelengths longer than 15\micron. This spatial overlap 
causes band mismatches of $\sim 10$-15\% at 
Channel 4; the band-matched spectrum has luminosity lower than
the pipeline luminosity by 6\%.
At $\lambda <15$\micron,  the spectrum is
better represented by the small aperture, and the D4 spectrum is distinct from that of D1. 

4) {\it Region D6, diffuse region to southwest}. Region center: 13$^h$ 39$^m$ 55\fs886, $-31^\circ$ 38\arcmin\ 26\farcs97; aperture radius: 0\farcs93. D6 is associated with the \mir\ continuum source
located 3\farcs 3 ($\sim 60$~pc) to the southwest of D1,
at 13$^h$ 39$^m$ 55\fs 88, $-31^\circ$ 38\arcmin\ 27\farcs 4 (1.3mm).
D6 has a diffuse  arclike appearance in ALMA CO images, and extended continuum
emission (Fig.~\ref{fig:regions}). The \hii\ 
region is visible as a pointlike source in the HST/NICMOS \paa\ image 
\citep{alonso-herrero2004}. 

5) {\it Ellipse, covering the visible \halpha\ and \paa\ plumes}. Region center: 13$^h$ 39$^m$ 55\fs952, $-31^\circ$ 38\arcmin\ 24\farcs42; aperture semi-major
axes: 2\farcs55 x 1\farcs53, p.a. 320.  Ellipse is centered on
D1, and encompasses that portion of the bipolar ``plumes"
of emission evident in visible continuum, \halpha, and \paa\ images that
fall on the \mrs\ chip \citep{calzetti1997,alonso-herrero2004}.  It includes
 the bright arcs to the northwest. 
We speculate that these plumes are emerging from holes in the
D1 cloud/cluster. The arclike shapes suggest flows rather than or in addition to 
simply escaping photons. Ellipse includes the regions
D1, Clusters 5 and 11, and D4. It excludes the regions D2 and D6.

6) {\it Chip, covering the footprint at 5\mic.} Region center: 13$^h$ 39$^m$ 55\fs952, $-31^\circ$ 38\arcmin\ 24\farcs42; aperture semi-major axes: 7\arcsec\ x 6\arcsec, p.a. 32. 
This region is a polygon slightly inside the borders of the 5\mic\
footprint, covering an area common to all MRS bands. 
Because of its irregular profile, the area of Chip is 41.6 square arcseconds.

7) {\it Chip minus D1, covering entire footprint at 5\mic\ minus the 0\farcs8 
region centered on D1.} This region is the Chip region minus the central region
around D1. It is defined as a separate polygonal region in
CARTA. Fringing mitigation is easier for spectra
from larger regions, preferable to subtracting the fringey D1
spectrum from Chip. There is some PSF contribution from D1 to
this region. 
Chip-D1 represents 
``everything except D1" fluxes for the 7\arcsec\ x 6\arcsec\ 
detector footprint.

The region of nitrogen enhancement localized by \citet{pruijt2026} does
not correspond to a distinct MIR continuum source. It is visible in nebular emission
lines, and appears to the north of region
D1. This region is also bright in high excitation fine structure lines, 
such as [Ne~III], and is discussed further in \citepalias{beck2025}, 
where we argue
that the photons illuminating this ``plumes" region arise in D1.

\subsection{Mid-Infrared Continuum Regions and Spectra}\label{subsec:res_spectra}

The starburst in \gal\ is of particular interest because it has
an exceedingly luminous cluster, $L\sim 10^9~ \rm L_\odot,$ 
within an exceedingly small volume, $r\lesssim 2$-3 pc. The radiation
field is intense and  relatively hard, with photons exceeding 40 eV and 
soft X-rays \citep{beck1996,  strickland1999, crowther1999, summers2004, allamandola1989, beck2025}. 
What are the effects of this radiation field on dust, particularly
small dust radiating in the mid-infrared? How far do these effects extend from the central
cluster, \gal-D1?  Since D1 and its neighboring regions all have the same
low metallicity in common, differences in their spectra may help
distinguish the effects that radiation field intensity and metallicity
play on their emission.

Spectra were extracted from the six regions with coordinates and sizes given
in the previous section. 
As discussed in $\S$\ref{sec:obs}, a pixel-by-pixel
spectral analysis directly from the cubes
is not possible because of spatial artifacts and 
strong spectral fringing due to the  bright D1 continuum source. 
Procedures for defringing spectra \citep{gasman2023}  work
best on 1D spectra rather than cubes. We chose the sizes of the regions according 
to the recommendation that spectra be extracted from regions with radius at least 
1.5-2 times the PSF FWHM \citep{law2023}. We adopted 0\farcs9, which is twice the 
PSF at 25\mic\ \citep{argyriou2023a} and corresponds to 16 pc at 3.7 Mpc.
For these apertures,
photometric errors due to various effects on the PSF should be $\sim 10$\%. 
Parts of D2 (and to a less extent, D6) overhang the edge of the chip at 5\mic,
although not significantly at longer wavelengths.
However D2 appears to be pointlike in the 5\mic\ images and we expect that
the flux we measure is accurate for this source. 

 Spectral region D4  requires care in interpretation because of its proximity to D1. D4 is 
difficult to see in the JWST continuum images. The D1 and D4 spectral regions overlap
somewhat,  but more critically, D4 lies within the point spread function ``petals" for $\lambda\gtrsim 12$~\mic. 
However, D4 is the most massive molecular cloud in the region, and our interest is in solid
state features in a cloud so close to D1. The features that we consider here arise at
wavelengths shorter than 15\mic, where the PSF contamination is $<10$\%. 
Fluxes for D4 at longer wavelengths are more contaminated by D1.
Details of the region sizes and centers are given in 
$\S$\ref{subsec:res_regions}. 

\begin{figure}[h]
\begin{center}
\includegraphics[width=\textwidth]{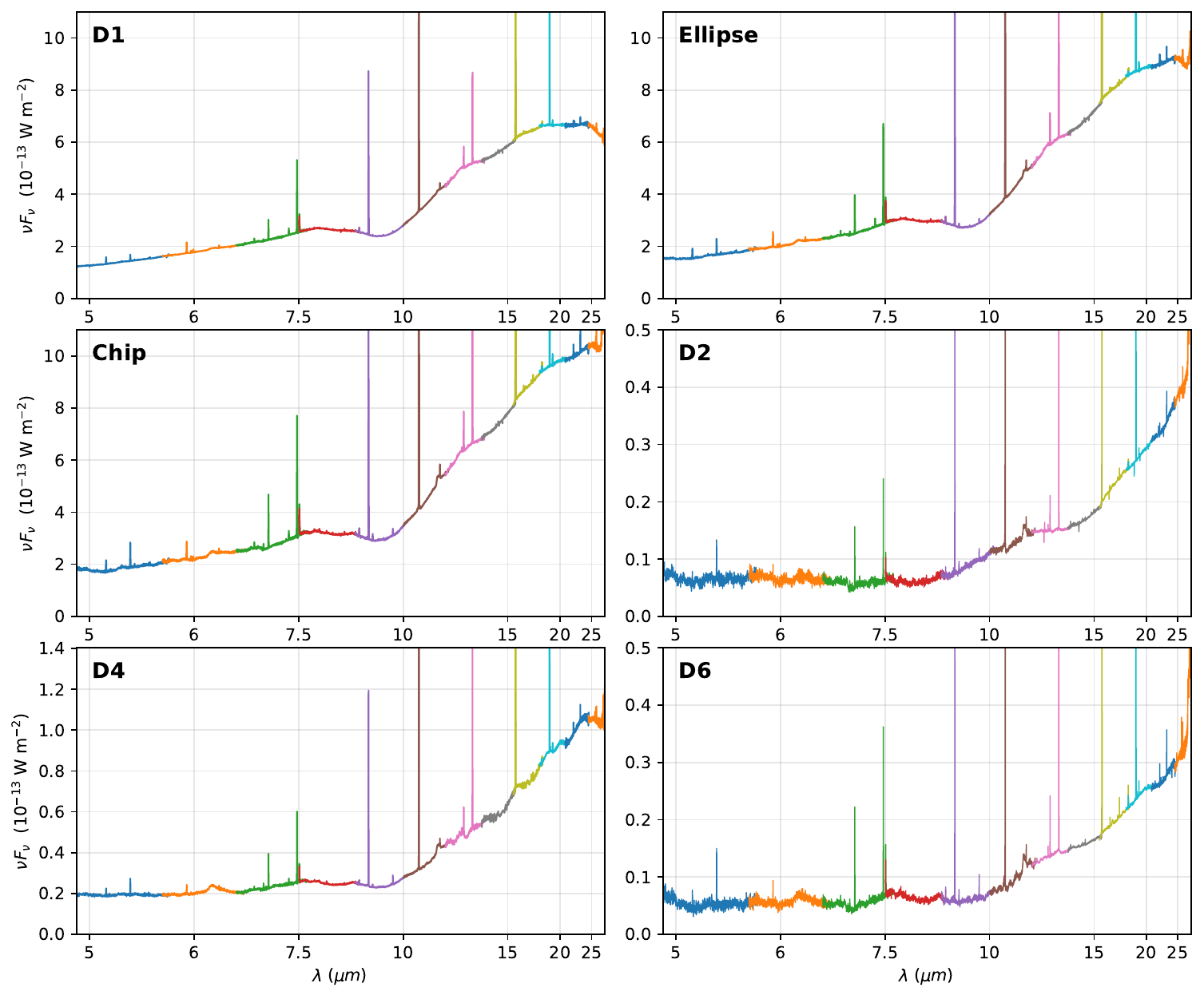}
\end{center}
\caption{ 
JWST \mrs\ continuum spectra for the spectral regions, from 1D defringed spectra.  
D1 dominates the larger Ellipse and Chip
regions, shown to the same scale. D1 shows a distinct 9.7$\mu$m silicate
absorption feature, as do Ellipse and Chip, the larger apertures that are dominated by D1. 
D2, D4, and D6 are far weaker. All regions show weak PAH emission. 
D2 has a weak 9.7$\mu$m silicate emission feature 
($\S$~\ref{subsec:res_silicate}).\\
The JWST \mrs\ continuum spectra are available as the data behind the figure.}
\label{fig:cspectra}
\end{figure}

The \mrs\ power spectra for the individual regions 
are shown in Figure \ref{fig:cspectra}.  In this paper we
focus on continuum features; lines are discussed in \citetalias{beck2025}. 
Color distinguishes the 12 spectral bands of \mrs.  Continuum emission is 
detected from 5\micron\ though  25\micron\ in all regions, where 
the detector response begins to degrade. 

The spectra have features in common with Galactic \hii\ regions.
  There is little evidence of emission from stellar photospheres, since
the spectra are flat or rising with wavelength. 
Emission at the shorter wavelengths has high characteristic temperatures, and 
is likely to be stochastically heated very small grains or continuum from PAH clusters. 
The D1 spectrum shows a significant increase with wavelength even
from 5-7.5\mic, which may indicate differences in the dust
population from the less luminous regions D2 or D6. Emission from classical larger grains
causes a rise in the spectrum beyond 10\mic.
The spectrum of D1,
as well as the Ellipse and Chip regions dominated by D1, 
 turn over at $\lambda\sim 20\mu$m, similar to 
the ISO spectrum of the Trapezium region in Orion \citep{cesarsky2000}. Degradation
of the chip performance at the longest wavelengths and the possibility
of broad, weak silicate absorption at 18\mic\ make it difficult to quantify this effect.

 On the other hand, the
spectra also differ significantly from Galactic \hii\ regions,  
 in part because of the low metallicity of \gal\ and in part due to the extreme
radiation fields of  supernebula D1. The main difference is that the 
 mid-infrared spectra of Galactic \hii\ regions 
such as Orion or W49 are dominated by broad line features from 
large molecular structures such as polycyclic aromatic hydrocarbons  (PAHs) \citep{cesarsky2000,stock2014,chown2023}. Solid state features have been long known to be weak in high radiation, high intensity, and low metallicity
environments such as \gal\ 
\citep[e.g.,][]{spoon2007, gordon2008a, engelbracht2008, hunt2010, lai2020, lai2025, zhang2025}.
PAH features are present in all the regions within \gal, 
including D1 where the radiation is most intense. 
However they are very weak everywhere, even in regions far from  D1.  We discuss PAH
emission beginning in $\S$\ref{subsec:res_PAHs}.  

The 9.7\mic\ silicate
solid-state feature is observed in some regions, but not all. 
Silicate absorption feature is clearly present in D1, as well as in the extended Chip and Ellipse
regions dominated by D1, as expected based on the  apparent high  extinction.  
We discuss the silicate features beginning in $\S$\ref{subsec:res_silicate}.

\subsection{Mid-Infrared Continuum Fluxes and Spectral Indices} \label{subsec:res_contfluxes}

In Table \ref{tab:cdetails} we present flux densities and mid-IR luminosities
at representative wavelengths for the regions. 
The wavelengths are chosen to be relatively free of PAH emission to 
facilitate comparisons with other \hii\ regions that have stronger PAH emission.  
Flux densities were measured from the 1D spectra for the
regions obtained from the MRS cubes using CARTA.
Values presented in Table~\ref{tab:cdetails} were interpolated from the 
loess-smoothed spectra to the specified wavelengths in R, as described in $\S$\ref{sec:obs}. 

\begin{deluxetable*}{lcccccccccccc}
\tabletypesize{\scriptsize}

\tablecaption{NGC~5253 Mid-infrared Continuum Properties}

\tablehead{
&
\multicolumn{5}{c}{Flux Density, $\rm S_\nu^b$} &
\multicolumn{4}{c}{Spectral Index, $S\propto \nu^\alpha$}
\cr                   \colhead{Region$^a$}&
\colhead{$\rm 2.8 \, mm$} &  
\colhead{$\rm 5.0 \mu$m}  & 
\colhead{$\rm 7.0 \mu$m}  & 
\colhead{$\rm 15.0 \mu$m}  & 
\colhead{$\rm 20.0 \mu$m}  & 
\colhead{$\alpha^{7}_5$} &
\colhead{$\alpha^{15}_5$} &
\colhead{$\alpha^{20}_5$} &
\colhead{$\rm log Q_0^c$} & 
\colhead{$\rm L_{OB}^d$} &
\colhead{$\rm L_{MIR}^e$} &
\colhead{$\rm L_{MIR}^{est~^f}$}
\cr                                                          &
\colhead{(mJy)} &    
\colhead{(Jy)} &  
\colhead{(Jy)} &   
\colhead{(Jy)} &   
\colhead{(Jy)} &  
&&&&
\colhead{($\rm 10^{7}\,\rm L_\odot$)}  &
\colhead{($\rm 10^{7}\,\rm L_\odot$)}  &
\colhead{($\rm 10^{7}\,\rm L_\odot$)}  
} 
\startdata
D1     &27.6   &0.20  &0.52   &2.9    &4.5   &-2.8   &-2.4  &-2.2  &52.69 &99   &33   & 51\\
D2     &0.85   &0.011 &0.013  &0.091  &0.20  &-0.4   &-2.2  &-2.0  &51.18 &3.0  &1.4  &2.2\\
D4     &1.2    &0.021 &0.041  &0.31   &0.62  &-2.0   &-2.5  &-2.4  &51.33 &4.3  &4.1  & 6.3\\
D6     &0.6    &0.010 &0.011  &0.083  &0.17  &-0.3   &-1.9  &-2.0  &51.03 & 2.1 &1.2  & 1.8\\
Ellipse&30.5   &0.22  &0.57   &3.6    &5.9   &-2.9   &-2.6  &-2.4  &52.74 & 110 &41   & 63 \\
Chip-D1&5.2    &0.10  &0.11   &1.2    &2.5   &-0.4   &-2.2  &-2.3  &51.97 &19   & 16  & 25\\
Chip   &32.0   &0.29  &0.61   &3.9    &6.6   &-2.2   &-2.4  &-2.3  &52.76 &115  &46   & 71\\
\enddata 
\tablecomments{
$^a$ Coordinates of aperture regions 
are given in $\S$\ref{subsec:res_regions}, and 
shown in Figure~\ref{fig:regions}.
$^b$Flux densities are interpolated to the specified wavelength from 
loess-smoothed spectra. Uncertainties are $\lesssim 5$\%, except 
 $\sim 10$\% for D2 and 
D6 at 5 and 7\mic, see $\S$\ref{sec:obs}.
Uncertainties in ALMA 2.8cm fluxes are $\sim 5$\%.
$^c$ Lyman continuum rates, $Q_0$,  are from 2.8~mm fluxes for
 $10^4$ K gas using $Q_{0} (s^{-1}) = 1.31 \times 10^{50} \nu_{11}^{0.118}~\rm
D_{Mpc}^2~ S_{100}$ (mJy). 
$^d~L_{OB}$ is the predicted O star
luminosity based on the $Q_0$  
using  $\log(L/L_\odot/Q_{50}/s^{-1}) = -43.7$ \citep{turner2017}, based on
Starburst99 models for a 1~Myr, z=0.008 cluster with upper mass cutoff
of 120\msun \citep{leitherer1999}.
$^e$Observed luminosity directly summed from line-free continuum MRS spectra.
$^f~L_{MIR}^{est}$ estimates the corrections to
observed luminosities to account for the Rayleigh-Jeans tail of the 
warm \mir\ dust, assuming peak wavelength of 20\mic\ and $\beta$=1.7.} 
\label{tab:cdetails}
\end{deluxetable*}

The total Chip fluxes can be compared to previous ground-based observations,
with similar apertures, 
as compiled in \citet{vanzisauvage2004}. Ground-based
fluxes at 12\mic\ range from 1.67 
to 1.9 Jy, with apertures of 4-7\arcsec; the JWST value is 2.3$\pm$0.2 Jy.
Some of this variation is due to filter differences. 
Fluxes at 20\mic\ range from 3.7-6.1 Jy for apertures of 6--8\arcsec,
respectively; 
the JWST value is 6.6$\pm$0.6~Jy.  The only direct comparison of ground-based
fluxes that can be made for the D1 region alone is 
to the high resolution ($\sim 0.3$\arcsec) Keck/Long Wavelength Spectrometer
images of \citet{gorjian2001}.
They measured 2.2 Jy at 11.7\micron, whereas JWST finds 1.7$\pm$0.2 Jy, and for
18.7\micron,  the MIRI flux is 3.6$\pm$0.4 Jy as compared to the 
Keck/LWS value of 2.9 Jy.
The fluxes are consistent within the uncertainties,
and demonstrate the importance of high  spatial resolution for the accurate 
characterization of extragalactic \hii\ regions.

To compare different sections of the \mir\ continuum, we have
computed spectral indices, $\alpha$ ($S\propto\nu^\alpha$). 
There is a distinct change in the continuum, particularly in
the less luminous D2 and D6, 
 such that the  emission spectrum below $\sim 10$~\mic\ is 
 significantly flatter than 
the rising spectrum beyond $\sim 10$~\micron. 
The dust spectrum below 10\mic\ is dominated by  hot and small grains,  possibly 
stochastically heated. Photospheric contributions have positive spectral index; 
there may be some small contribution at the shorter wavelengths 
in D2 and D6 that flattens the 5-7\micron\ continuum, but photospheres
contribute negligibly in this region
compared to hot interstellar dust emission in D1.   
The longer wavelength radiation tends to
be  dust in equilibrium heated by intense radiation fields by the cluster stars, 
although large nanoparticles with $10^3$-$10^4$ C atoms that
are stochastically heated can also in theory
produce radiation even at these wavelengths \citep{draine2021}. 
To quantify the  spectrum 
we compute spectral indices, $\alpha$,  from flux densities 
at wavelengths of 5\micron, 7\micron, 15\micron, and
 20\micron, wavelengths relatively free of PAH emission. 
These spectral indices are listed in Table 1.
 
The spectrum of D1 between  5 and 15 or 20\mic~has 
$\alpha\sim~$-2.2 to -2.4, only slightly steeper than the less luminous \hii\ regions 
D2 and D6, which have ratios of -1.9 to -2.2.  The Trapezium region of
Orion \citep{cesarsky2000} has index similar to or slightly flatter than D1, with
$\alpha\sim~$-1.8 to -2.3. 

The bigger difference is at the shorter wavelengths, where the
spectral index of D1 differs
significantly from D2 or D6. 
The D1 spectrum rises  from 5\mic\ to 7\mic\ with spectral index -2.8, compared 
to -0.3 to -0.4 for regions D2 and D6.  Unlike D1, the spectral indices of D2 and D6 steepen significantly  beyond the silicate feature. 
This steepness
in the D1 5-7\micron\ region could be due to the onset of a plateau
near 7-9\micron.

We can estimate an interstellar 
radiation field from the spectrum in terms of U, the ratio of
the field to the standard interstellar radiation field of
\citet{mathis1983}. Based on the models of \citet{draineli2007},
this spectral index  and the turnover of the dust spectrum at $\sim 20$\mic,
suggest that the interstellar radiation field  in D1 has $U\sim 10^{5.5}.$ 
This is close to the radiation field in which the UV photon fluxes become 
high enough that the grain heating of small dust particles
is no longer stochastic \citep{draineli2007}. 

\subsection{The Ratio of Mid-infrared to Radio Continuum Flux} \label{subsec:res_MIRradio}

Free-free fluxes are also listed in Table~\ref{tab:cdetails} for
the spectral regions. These were measured  for the individual 
regions from 107 GHz (2.8 mm) ALMA maps of 0\farcs5 resolution
(ALMA Project 2013.1.00833.S, P.I. E. 
 Rosolovsky). At $\lambda = 2.8$~mm, the emission is
 free-free emission, with negligible contributions from dust
or synchroton emission. The largest angular scale of these images
is 4\arcsec, which is close to the size of the Chip region, so missing flux
from short baselines is negligible. 
From the free-free flux we derive extinction-free Lyman continuum
rates, which are also given in Table~\ref{tab:cdetails}. The 
 extinction in the \mir\ is not insignificant for D1; this
is discussed further in \citetalias{beck2025}.

The total Lyman continuum rate for the D1 \super\ 
core is \logq\ = 52.68, while the entire Chip region has \logq = 52.76  
\citep{mohan2001,rodriguez-rico2007,bendo2017}.
The \nlyc\ value for the Chip region is consistent with that
predicted by VLA and ALMA images \citep{turner1998,rodriguez-rico2007,bendo2017}.

The continuum spectra allow us to compute the ratio of \mir\ to free-free
flux for the \hii\ regions in \gal. The excellent spatial correlation of free-free
and \mir\ dust emission within Galactic \hii\ regions 
has long been known \citep[e.g.,][]{genzel1982}.
The predictable relation of \mir\ and free-free flux is 
an important diagnostic of O star formation, and 
this relation has been used previously in
\gal\ to argue that the luminous D1 is most likely an \hii\ region and not
an AGN \citep{gorjian1996}.  The high spatial resolution
of JWST allows individual \hii\ regions to be isolated so that 
the relation can be studied in other galaxies.

We obtain ratios of $S_{20\mu m}/S_{2.8mm} = 160$, 230, and 280 for the
isolated \hii\ regions D1, D2, and D6, respectively, and 190-200 for
the extended regions, Ellipse and Chip. 
This can be compared to the values seen in W51 
of $S_{20\mu m}/S_{6cm}\sim 100$ \citep{genzel1982},
which, translated to 2.8mm, would be a ratio of 140.  The agreement
between the Galactic \hii\ value and that measured in D1 is
excellent, because D1 is more nearly resolved. The smaller D2 and D6
\hii\ regions have more contribution from molecular clouds.

\subsection{Mid-Infrared Luminosities and Photon Escape in NGC~5253} \label{subsec:res_escape}
Observed \mir\ luminosities, \lmir, 
are computed from the observed spectra and are listed in 
Table~\ref{tab:cdetails}. These fluxes correspond only to the measured
fluxes in the 4.9-28\mic\ spectral range, not beyond, and thus are an
underestimate of the total IR luminosity emitted by ``warm" dust. 
We correct for the Rayleigh-Jeans tail of the distribution of the flux represented
here by modeling the spectra as  modified blackbodies with peak
emission up to 20\mic, as indicated by the spectra (Fig.~\ref{fig:cspectra}) 
and is similar to the peak in Orion \citep{cesarsky2000}. 
We assume an emissivity $\beta\sim \nu^{1.7}$. For these parameters the Rayleigh-Jeans
 tail contributes $\sim$35\%. The extrapolated blackbody luminosities are
listed as $L_{MIR}^{est}$; these are still underestimates since they
do not include the power by dust with peak emission at longer wavelengths,
the ``cold" dust \citep[e.g.][]{remy-ruyer2013,remy-ruyer2015}.  For D1,
Ellipse, and Chip, the spectra show turnovers at 20-25\micron, so $L_{MIR}^{est}$
is probably fairly accurate for the region. 
From the Lyman continuum rates, Q$_0$, we predict a corresponding 
luminosity due to star formation, designated as $L_{OB}$ 
based on STARBURST99 models (Z=$0.25Z_\odot$, mass
cutoffs 1 and 150 \msun, continuous star formation, age=1 Myr). 

There are two potential corrections to be made to the radio 
fluxes. Radio free-free emission is affected by dust, since the
dust directly competes with gas for ionizing photons. It
is possible to correct for the effects of direct absorption of UV by dust, 
accounting for the fraction of dust heated by Lyman $\alpha$ photons, as described
by Inoue \citep{inoue2001,inouehirashitakamaya2001}.  Using their
curve for the LMC for average dust absorption efficiency, $\epsilon$,
we adopt a value of unity, 
since $A_V$ is high. From the values in Table \ref{tab:cdetails},
using \lmirest\ for the luminosity, we obtain a fraction, $f$, of
ionizing to total UV photons of $f\sim 1.06$ for D1.  
In other words, there is negligible
direct absorption of UV photons by dust within the D1 \hii\ region. 
The computation changes little, $f=1.01$, for the values for the entire Chip region.
For comparison, typical values for Galactic \hii\ regions are $f\sim 0.3$
\citep{inouehirashitakamaya2001}. Thus we do not correct for
direct absorption of UV by dust within the D1 \hii\ region.

The second potential correction to the radio flux is for 
UV photons escaping beyond D1 and beyond Chip. 
Comparing the free-free fluxes of D1 and Chip suggests that most of the ionizing
flux within the central 100 pc region has its origin in D1.  Porosity has
been suggested to explain extended [CII] emission from low metallicity galaxies 
\citep[e.g.,][]{cormier2019},
although the bright CO emission \citep{turner2015, turner2017, consiglio2017} 
and high $A_V$ of D1 are less typical of low metallicity environments. 
The free-free flux from D1 is $\sim90$\% of the total free-free
flux within Ellipse, and 83\% of the flux within Chip,
which contains other \hii\ regions.  The Ellipse region has
visible plumes of nebular emission, as can be seen in
Figure~\ref{fig:7micron4panel}. Beyond the 100 pc region 
covered by Chip, VLA continuum observations
at 33 GHz \citep{rodriguez-rico2007} 
predict a flux at 107 GHz of $\sim 40$-42~mJy, as compared with
32 mJy observed within the Chip region. If these
photons arise within D1, then $\sim$25\% of photons escape beyond the
\mrs\ chip. 

We  propose, based on the comparison of the luminosities implied
by the mid-IR emission and the radio free-free fluxes, that $\sim 25$\%, of the UV 
photons are escaping from the $r\sim 15$pc D1 aperture, even though
the apparent extinction along the line of sight is high ($A_v\sim 35$). 
Photons can escape if the D1 cluster/cloud is porous. 
These escaping photons are absorbed by dust beyond a radius of 50-60 pc from D1.
Additional evidence that D1 is the source of extended nebulosity is presented
in \citetalias{beck2025}.

\subsection{Silicate Feature} \label{subsec:res_silicate}

\gal\ has a relatively low metallicity, $Z\sim 0.3~Z_\odot$, and an incredibly 
luminous young cluster. Both factors can create an environment hostile to solid particles or large molecules, including large hydrocarbon molecules such as PAHs. 
On the other hand, \gal\  has high visual extinctions and dense molecular clouds, 
including molecular gas within the $r<3$pc core of D1 \citep{turner2017, consiglio2017}, indicating that dust and large molecules exist close to and even within
the great cluster. What do the current observations tell us about solid state features in \gal ? 

 Solid state features are  apparent in the spectra of all regions as shown in Figure \ref{fig:cspectra}. These include
   silicate features, in both absorption and emission, and emission
  from large carbon mega-molecules, which we refer to here as polycyclic aromatic hydrocarbons (PAHs) for simplicity, although not all of these hydrocarbons are aromatic. 
In this section we characterize the silicate and PAH features 
in \gal\ to find their strengths, equivalent widths, profiles, 
and wavelengths.  These data can reveal the effects of different radiation fields on a host environment of uniform metallicity.  We discuss first the silicate features and then the PAHs.  

\subsubsection{Silicate Feature: Absorption and Emission}\label{subsubsec:res_silicate_abs_em}

\begin{figure}
\begin{center}
\includegraphics[width=5.5in]{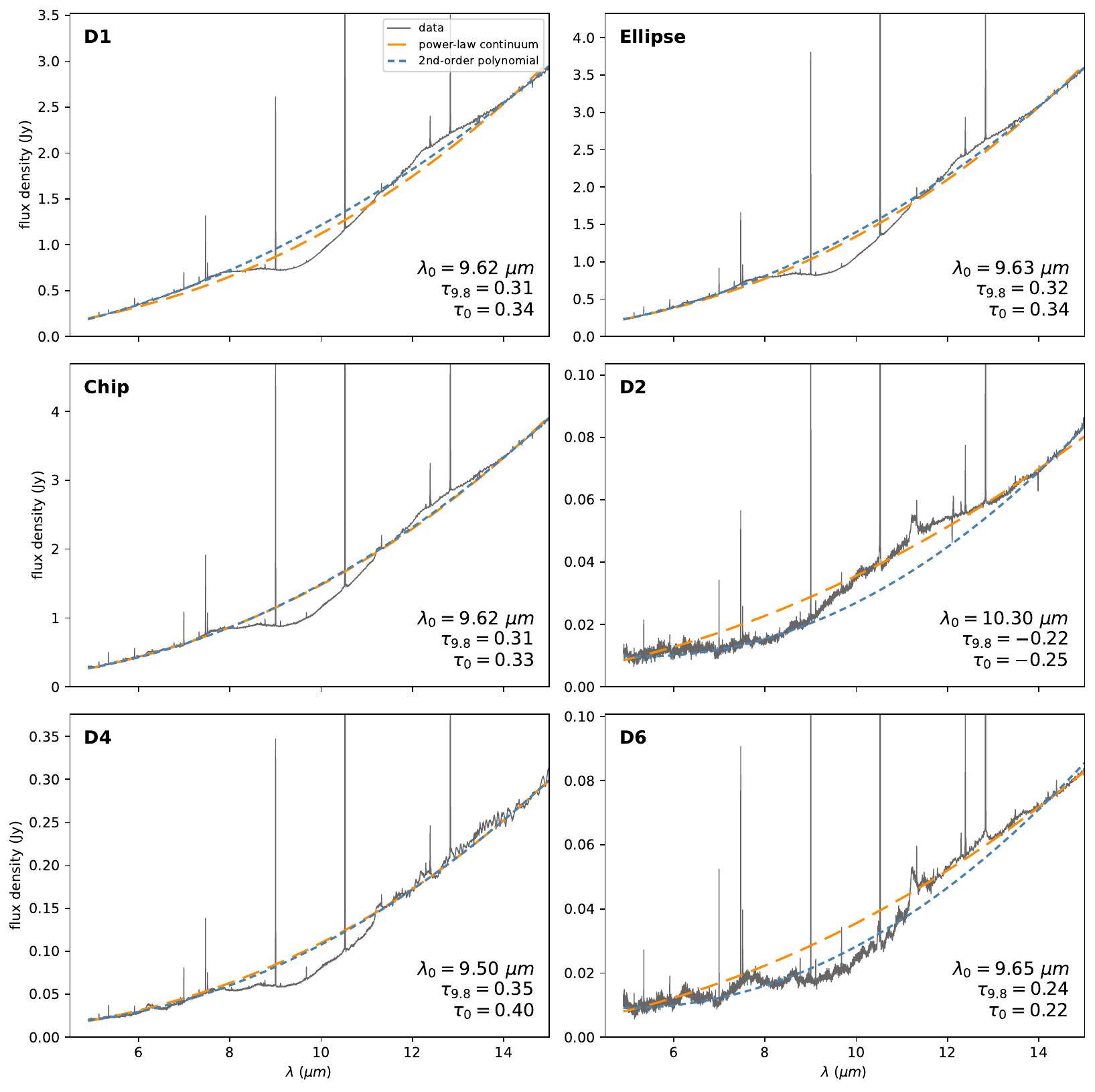}
\end{center}
\caption{
Spectra from 5 to 15 \mic\ with fits to the continuum underlying 
the 9.7\mic\ silicate feature overplotted. Shown are second order polynomial and power law fits 
to the continuum from 5.0-5.5\micron,  6.97-7.2\micron\ (polynomial only), and 14.3-14.5\micron,  
 which were used to fit the silicate feature. Also shown are 
 the peak wavelengths for the Drude profile fits to the
 silicate feature for each region, and 
 the optical depths at the profile peaks
 and at 9.8\micron\ based on the quadratic continuum.
}
\label{fig:silicate}
\end{figure}

The  broad feature centered at $\sim 9.7$\micron\ is 
usually associated with amorphous astronomical silicates \citep{whittet1992,dorschner1995,kemper2004}; recent evidence  \citep{gordon2023} supports this idea.  The silicate feature is detected in \gal\ for the first time here.
Plots of the 5-15\micron\ spectrum for every region can be seen in Figure~\ref{fig:silicate}. 
The 9.7\mic\ feature is strongest in D1 and is seen as well in the 
Ellipse and Chip regions, which are dominated by emission from D1. 

\citet{spoon2007} calculated optical depths for the silicate feature by 
fitting a power law continuum 
between 5.5\mic\ and 14.5\mic\ and from that predicting
a 9.8\micron\ value in the absence of a silicate feature; \tausil\ is defined 
as the natural log of the ratio of observed to predicted flux. 
We followed this procedure, and also fit a second order
polynomial as an alternative 
before fitting modified Drude
profiles to the resulting silicate features, as described
in the Appendix. Following the suggestion
of \citet{brandl2006} we extended the fit to 18-23\mic, which required a fourth order polynomial. While the fourth order polynomial is a very good fit overall to the data out to 25\mic\ for D1 
(Fig.~\ref{fig:Silicate2panel}), it is not as good for 5-15\micron\ continuum. 

Data and second order polynomial and
power law fits to the continuum  are
shown in Figure~\ref{fig:silicate}. Also shown are the peak wavelengths, 
$\lambda_0$, fit for the Drude profiles, and the resulting
optical depths at $\lambda_0$ and 9.8\micron\ based on the second
order polynomial continua. Uncertainties in $\tau_{9.8}$
due to the difference in continuum fits are 0.02-0.08
for most regions, with the exception of D6, at 0.2, and D2
discussed below. Profiles of the silicate features obtained by
subtracting the data from the 2nd order and power law continua are shown for sources D1 and Chip in
Figure~\ref{fig:Silicate2panel}
 
Region D1,  and the Ellipse and Chip regions dominated by D1, each show
a clear 9.7\mic\ feature.
The strength of the silicate feature is sensitive to choice of
smooth continuum fit, and is complicated by the possible presence of 
excess flux near 12.2\micron, discussed in the next section.  We fit
the silicate feature for the limited range 7.8 to 11.9\micron.  
The power law fit specified by \citet{spoon2007} gives \tausil = 0.23;
the second order polynomial value is \tausil\ = 0.31; we adopt the weighted mean value, 
which gives $\tau_{9.7}=0.27\pm 0.04$  for D1. 

The silicate feature is weaker than would be predicted from the observed visual extinction 
based on Galactic observations. The visual extinction 
of $A_V \sim 30$-35 is derived from
comparison of HI recombination lines and VLA radio continuum, most recently with MRS observations \citep{calzetti1997,beck2025}. 
(While we use the common term ``visual extinction", these are pc-scale regions rather than sightlines 
to stars and ``attenuation" may a more appropriate technical description, 
although as these are isolated \hii\ regions  mixed extinction models are not applicable.)
In the Galaxy, the optical depth of the 9.7\mic\ silicate feature  
is related to visual extinction by the empirical relation 
$A_V/\tau_{9.7} = 18.4 \pm 1.0$ \citep{whittet1992}. For D1 the ratio is $A_V/\tau_{9.7} \sim 130 \pm 20$, higher than
the Galactic prediction by factors of 5-10. However, 
the diffuse cloud prediction does not appear to hold in denser Galactic
regions such as Serpens or Taurus \citep{chiar2007}, which may be environments more similar in molecular gas properties to D1.

In D2 the silicate feature appears in emission in
in the quadratic continuum model, and 
consistent with $\tau_{9.8}\sim0$
for the power law continuum. 
D2 has significant extinction of  $A_V \sim 15$ \citepalias{beck2025},
from which the Galactic relation would predict a silicate optical depth near unity, which would be detectable. 
Silicate emission is typically observed toward evolved stars; 
 the D2 spectrum in this region is reminiscent of
the ``big-11" feature attributed to SiC in AGB stars
\citep{sloan2014}. 
Silicate emission has also been observed in \hii\ regions,
including the Trapezium region in Orion, in NGC~3603, and in the SMC sources NGC~346 and N66B 
\citep{cesarsky2000,lebouteiller2007,whelan2013}. 
\citet{whelan2013} 
suggest that it indicates a very brief phase of cluster evolution.  D2 has a visible \hii\ region
and also a closely associated molecular cloud, thus it is likely of moderate age and older than D1.  

A silicate feature at 18.7\mic\ is observed in
Galactic sources \citep{gibb2004}. It is less deep than the 9.7\mic\ and very broad \citep{chiar2006}. 
We do not see clear evidence of the 18.7\mic\ feature in the spectra of \gal, but this is not
surprising given that it would be half to one third
the optical depth of the 9.7\mic\ feature. It is also not apparent
in the Spitzer spectra of \citet{beirao2006}.

\subsubsection{Silicate Feature Profile in D1}\label{subsubsec:res_silicateprofile}

\begin{figure}
\begin{center}
\includegraphics[height=3.5in]{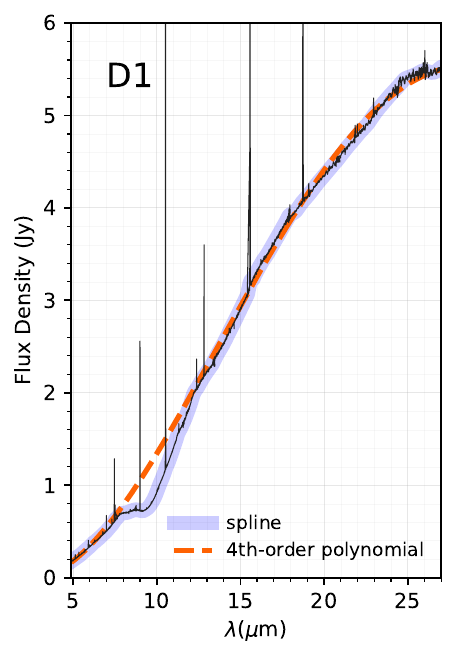}~~~
\includegraphics[height=3.5in]{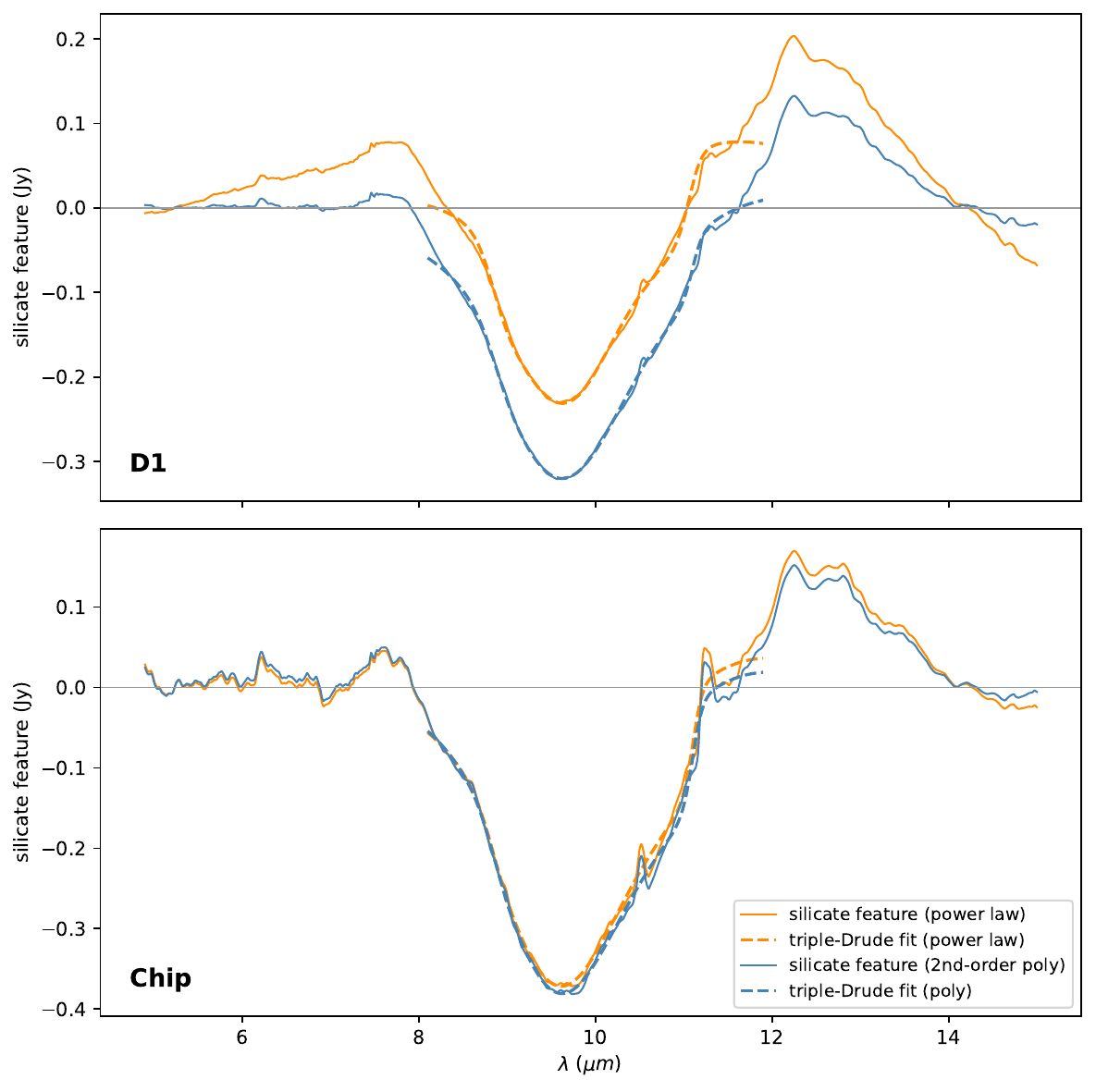}
\end{center}
\caption{Continuum fits and corresponding silicate features in region D1.
{\it{(Left}:}
D1 flux is shown with a fourth order polynomial fit to the full 5-28\mic\ MRS
spectrum (dashed line). Also shown is the spline fit to the continuum
using knots similar to those used for the Orion PDR spectrum of \citet{chown2023}.
{\it(Right):} Two models for the silicate feature, constructed from the difference
between the line-free loess fit to the data and the 2nd order polynomial fit to
 the continuum (orange lines) and the power law fit (blue lines) 
for regions D1 and Chip. Modified Drude functions were simultaneously fit to the 
silicate feature and the 8.6 and 11.2\mic\ PAH features between
8.3 and 11.9\micron\ as described in the text; the Drude fits to the silicate feature
is shown as dashed lines for the two continuum models. 
}
\label{fig:Silicate2panel}
\end{figure}

In D1 the silicate feature is strong enough that we can model its  profile and compare to Galactic sources. We fit the silicate feature to a modifed Drude function, following \citet{gordon2021}. 
 Because the silicate feature extends to include  PAH emission at
 8.6 and 11.2\micron,  we fit a triple Drude model with profile parameters for both silicate and 
 PAH components (see Appendix). Double Drude fits for the features corresponding to two underlying continuum options, second order polynomial and power law, are shown in Figure \ref{fig:Silicate2panel}b.

Values fit for the modified Drude parameters for D1 are  
$\lambda_0 = 9.622 \pm 0.006$ (rest wavelength), $a = 0.28\pm 0.05, ~\gamma_0 = 2.05\pm 0.05$. Values for Ellipse and Chip are the same to within the uncertainties. 
These values are within the range of parameters found
by \citet{gordon2021} for their single Drude fits of the silicate feature in
diffuse Galactic gas, $\lambda_0$ (9.64 to 9.94), $\gamma_0$ (1.52 to 3.99),
and $a$ (0.07 to -1.25). \citet{gordon2021} find that $\lambda_0$ is correlated with 
both $\gamma_0$ and $a$.  Compared to Galactic regions, the D1 value of
$\lambda_0$ is on the short wavelength side of the observed
range,  and $\gamma_0$ is as predicted by their relation. 
The asymmetry parameter, 
$a$ is smaller than predicted by their relation, given $\lambda_0$. Thus the
silicate feature profile is similar to those  observed in Galactic spectra.

\subsection{PAH Features: Fluxes, Profiles, Peak Wavelengths}\label{subsec:res_PAHs}

In addition to the silicate absorption feature, there
is emission from organic macromolecules, often, although
not always, polycyclic aromatic hydrocarbons, and
we will refer to all of these as ``PAHs". PAH emission appears in the near- and mid-infrared spectra of a wide range of astronomical sources. Major PAH emission features occur at 3.3, 6.2, 7.7, 8.6, 11.2 and 12.7 \micron.  Spectral classes A, B, C, and D have been established based on the 6-9\micron\ and 11-14\micron\ emission {\citep{peeters2002, matsuura2014}}: Class A is associated with the ISM, OB stars and \hii\ regions, classes B and C with young stellar objects, disks, T Tauri and Herbig Ae/Be stars, and Class D with carbon-rich post AGB stars. 

The PAH features in \gal\ 
 are very weak compared to the spectra of more metal-rich
 galaxies \citep{brandl2006,smith2007}.
Low metallicity and intense radiation fields weaken or suppress PAH emission 
\citep{jackson2006, madden2006, wu2006, smith2007, engelbracht2008, gordon2008a, galliano2008b, hunt2010, lai2024}. 
Both conditions obtain in \gal.  As all regions in this
study lie within a radius of $\sim 50$ pc and will have had similar initial metallicity; differences in their spectra will be due to local conditions.  A complete modeling of all features  \citep[e.g.,][]{smith2007, maragkoudakis2026}  requires the  key 3\micron\ features \citep{lai2025,maragkoudakis2008} 
not in this data set.
 However, since these \mrs\ observations isolate individual \hii\ regions, with a wide variation in environment, 
examining the individual 5-15\micron\ spectra may bring insights into the superluminous D1 environment.

The 6.2, 6.8, 7.7, 8.6, 
and 11.3 \micron\ features are observed in all regions within the central
starburst in \gal.  Minor features
 at 5.3, 6.1, 7.2, 9.3, 11.0, and 12.7 \micron\ are also present.
We isolate individual features by subtracting from the spectrum a continuum obtained from a cubic spline fit to
the line-free loess-smoothed line-free spectra.  The
resulting features were fit by modified Drude profiles and Gaussians
within specific wavelength regions as described in the Appendix.  
Weak features in the 5\micron\ band, including 
5.7\micron,  appear in the spectra of D4 and Chip-D1 but were not fit,
since they are complicated by the presence of many series 7 HI lines \citepalias{beck2025}.

The continuum-subtracted features are shown in Figure~\ref{fig:pah_plots}
with their peak rest wavelengths, $\lambda_0,$ from the modified Drude fits. 
For strongly asymmetric features such as those at 6.2 and 11.2\micron\, the
Drude wavelengths are shifted by 0.01-0.03\micron\ with respect to peaks for Gaussian fits. Both Gaussian and Drude fits, and their sums, are shown.

Fluxes and equivalent widths for the brighter features, as determined from Drude function fits, are given in Table~\ref{tab:pahs}.  Based on variation among different spline fits described
in the Appendix, fluxes for the 
6.2, 6.8, and 11.2 
 features are estimated to be good to better than 20-30\%, 
with the exception of 11.2\micron\ in D1, which is 35-35\%. 
 Since the spline fit was done in an identical manner for all regions, 
intercomparisons among regions can be made. 

Because their equivalent widths are so small, and 
because they are so dependent on the spline fit 
in the neighborhood of the broad silicate feature, fluxes of 
features between 7--9\micron\ (7.2, 7.6, 7.8, 8.6\micron) 
are uncertain to a factor of two 
and possibly more in D2 and D6, with 8.6\micron\ particularly
sensitive to the spline. Thus we report only the
sum of the features from 7-9\micron, rather than the individual
features. Based on spline fits with different knot choices, we
estimate that the uncertainties in the summed fluxes are
$\lesssim 30$\%, while fluxes of individual features can vary
based on knot choice within the 7-9\micron\ region 
by up to a factor of three.

\subsubsection{6 Micron}\label{subsubsec:pahs_six}
\begin{figure}
\begin{center}
\includegraphics[width=\textwidth]{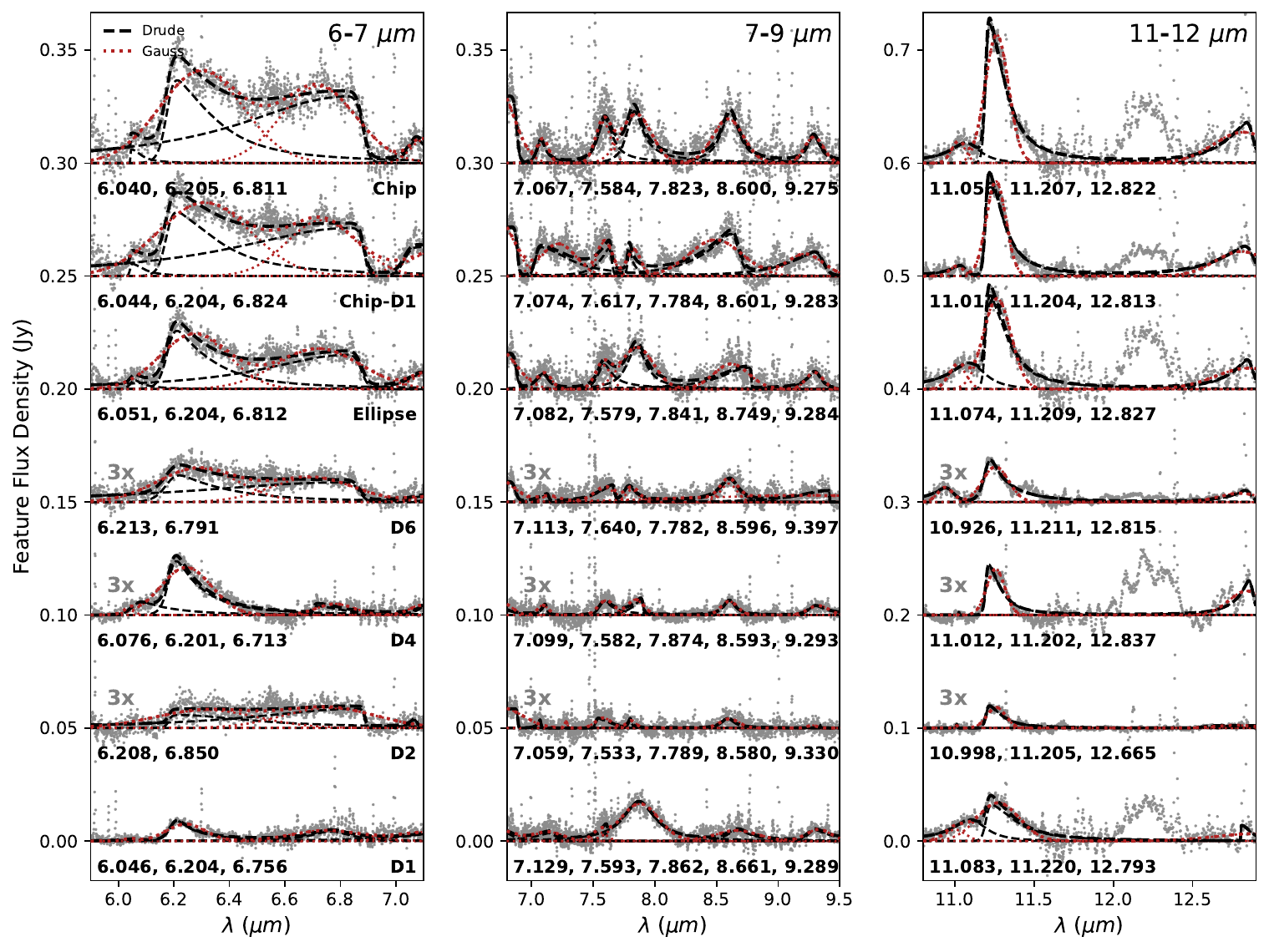}
\end{center}
\caption{
Continuum-subtracted PAH features. 
Regions are ordered vertically as identified in the left panel 
and plotted such that the zero points are 
offset by multiples of 0.05, 0.05, and 0.1 Jy, left to right. 
Spectra for regions D2, D4, and D6 have been scaled by a factor of three. 
Data are presented as a scatter plots in grey, and include nebular lines. Modified Drude
profile fits are shown in dashed black lines; Gaussian
fits in dashed red. Heavy lines are sums of the Drude (black) and Gaussian (red) fits. 
Rest peak wavelengths, $\lambda_0$, from the Drude fits are
given, in microns. The Drude peak can differ by 0.04-0.1\micron\ 
from the Gaussian fitted peak, with larger values for the more asymmetric 
features such as those at 6.2 and 11.2\micron. 
{\it (Left):} 6.2~\mic\ features. 
{\it(Center):} 7-9~\mic\ features. 
{\it(Right):}  11-12~\micron\ features. 
The 12.2\micron\ feature is an artifact. 
Continuum subtraction and profile fits are described in the Appendix. 
 }
\label{fig:pah_plots}
\end{figure}

The 6\micron\ features  attributed to C-C stretching modes \citep{tielens2008}
 are clearly detected as shown in Figure~{\ref{fig:pah_plots}}a.  In 
 \gal\ the 6\micron\ feature in 
\gal\ is very broad, extending from 6.2 to nearly to 6.9\micron, combining 
two features at $\sim$6.2 and $\sim$6.8\micron. The 6.2\micron\ feature rises steeply on 
the blue side, and is bookended by a steep red edge past 6.8\micron.  The combination of two Drude 
profiles gives a good fit to the combined feature profile, which extends from 6.15-6.85\micron.  

The 6.2\micron\ profiles are 
asymmetric and are well fit by modified Drude functions 
with negative asymmetry factors $a<0$. They are not as well fit by Gaussians. 
 The steep blue rise and peak wavelengths of 6.20-6.22 are characteristic
of class A sources, which originate in or near \hii\ 
regions \citep{peeters2002}.  This feature is likely due to PAHs with nitrogen-substitutions, or PANHs \citep{peeters2002, hudgins2005, ricca2021}.

The emission at 6.06\micron, which is detected in
Ellipse, Chip, and Chip-D1,  is strongest in Chip-D1, 
the extended region outside D1. This feature 
is more closely associated with class B sources, 
such as PNe, post-ABG objects and Herbig Ae/Be stars \citep{tielens2008}. 

There is a hint of a feature at
6.5\micron\ in regions except D4 and possibly D1. 
This feature could be due
to smaller hydrocarbon PAHs \citep{hudgins2005, bauschlicher2008} or fullerenes \citep{jensen2022}. This is near the band overlap region, which extends from
6.53-6.63\micron. We have tried the spline fit without spectrum
stitching and the results are nearly the same, with fit amplitudes 
varying by $\sim 5$-6\%, so this is unlikely due to band
matching and could be a real feature. 

The 6.8\micron\ feature persists over all regions.
Its sharp red cutoff does not correspond to any known spectral
artifacts.  The feature is strongest in the larger regions, Chip and especially Chip-D1; since Chip-D1 is nearly as strong, the 6.8\micron\
feature is largely coming from the region outside D1.  A feature at
6.9\micron\ feature  originates in  aliphatic (non-aromatic) structures 
\citep{bauschlicher2008}; we discuss in $\S$~\ref{sec:discussion} 
the implications of aliphatic emission in \gal.

\subsubsection{7-9 Micron}\label{subsubsec:pahs_sevennine}

The 7-9\micron\ features in supernebula D1  are overall 
similar in intensity to those at 6.2\micron, as expected for active star-forming regions \citep{galliano2008b}.
These features, attributed to stretching, bending and wagging modes of  CC and CH \citep{tielens2008}, are shown in Figure~{\ref{fig:pah_plots}}b.  This spectral region has many weak features in addition to the stronger 
7.6, 7.8 and 8.6\micron\ features, and also broad continuum plateaux   \citep{peeters2002,peeters2004,brandl2006, tielens2008, andrews2015, galliano2022, hrodmarsson2025}. 

The fluxes of features at 7-9\micron\ are 
particularly sensitive 
 to the spline continuum fit, since an underlying continuum
 plateau and onset of silicate absorption occur in this spectral region.
 An additional complication is the proximity of band overlaps at 
 7.51-7.65\micron\ and 8.67-8.77\micron, which can affect these
 weak features. 
 The ratios of the 7.7\micron/8.6\micron\ features are 
 quite sensitive to the spline fit,
 and there is sparse a priori information on the continuum in this
 region, which can vary significantly due to the grain size 
 distribution \citep{galliano2022}.  Only in features very  
 close in wavelength, such as the 7.6 and 7.8 features, can
 intensities reliably be compared. Peak 
 wavelengths are also relatively robust to spline fit. 

Near 7.7\micron\ two distinct components can be identified and fit in all regions: the one at
7.58\micron, which is an aromatic C-C stretch and another at 7.86\micron, which corresponds to a 
C-C stretch+ C-H bending \citep{draineli2007}. The 7.86\micron\ feature is dominant in D1, 
and to a lesser extent, in the Ellipse and Chip regions containing D1.  The 7.58\micron\ feature is more prominent 
in Chip-D1 and also D4, regions that contain more neutral or cooler molecular gas. 
The combined fluxes of these two features represent nearly
 half of the total PAH flux in D1, and 25-30\% in the other regions. 

Weak features at 7.1\micron\ and 9.3\micron\ are seen. The 9.3\micron\
 feature does not correspond to any PAH feature that we are aware of. The
 7.1\micron\ feature is strongest in Chip-D1.

 \subsubsection{11-12 Micron}\label{subsubsec:pahs_eleventwelve}

The 11\micron\ features, due to out of plane bending
modes of C-H in neutral aromatic hydrocarbons \citep{hony2001, 
vanDiedenhoven2004, tielens2008}, are  shown in Figure~{\ref{fig:pah_plots}}c. 
 The 11.2\micron\ feature is detected 
in all regions.
Pronounced steep blue edges are seen; modified Drude profiles 
have asymmetries of $a=-29$ to -32. 
 
The 11.0\micron\ feature is also detected. It 
is remarkably strong in D1 and to a lesser extent, in   
regions containing D1 
(Ellipse and Chip). It is much fainter relative to the 11.2\micron\ feature in the other regions.
 In the Galaxy, the 11.0\micron\ wing is stronger in Class $\rm B_{11}$ sources
\citep{hony2001,vanDiedenhoven2004,chown2023}, which are  associated with circumstellar
material in post-AGB stars, PNe and Herbig Ae/Be stars \citep{tielens2008}. 
In the relative strength of its 11.0\micron\ emission, supernebula D1 has a distinct $B_{11}$ character. 

A weak unidentified feature  at 11.54\micron\ was detected in D1,
Ellipse, Chip-D1 and Chip regions,
 but was not fit well by a Drude profile. This feature coincides with
the HI 15-9 line, but is far broader than observed for HI lines in \gal\  \citepalias[][]{beck2025}. 

The PAH feature at 12.8\micron\ that can be nearly as strong as the
11.2\micron\ feature 
\citep{hony2001} appears to be present. It coincides with
the [Ne II]12.8\micron\ line, which is much narrower and has
been weighted out of the analysis.  
This feature therefore appears to be real but
the flux may be affected by the wings of the neon line. 

The feature appearing at 12.2\micron\ is a 
known artifact due to a leak from another band.  The leak transfers 
2.5\% of the 6.11\micron\ flux \citep{argyriou2023a} to
a feature at 12.2\micron\ with a width of $\sim$1\micron. For  D1 this corresponds to  
$F_\nu \sim$0.01 Jy, close to what is observed.  

\begin{deluxetable*}{lcccccccccccccccc} 
\tabletypesize{\footnotesize}
\tablewidth{0pt}
\setlength{\tabcolsep}{2pt}

\tablecaption{NGC~5253 Broad Mid-Infrared Spectral Features}

\tablehead{
\colhead{$\rm Region$} &
\multicolumn{7}{c}{~~~\hrulefill~~ Flux$^a$~~\hrulefill~} &
\multicolumn{5}{c}{~\hrulefill~~ EW$^a$~~\hrulefill~} &
\colhead{~$\rm \frac{F_6}{F_{11}}^b$} &
\colhead{$\rm \frac{PAH}{7}^c$} &
\colhead{$\rm \frac{PAH}{14}^c$} &
\colhead{$f_{C}^{d}$} 
\cr
&\colhead{7$\mu$m$^c$}&  \colhead{14$\mu$m$^c$}& \colhead{5.3$\mu$m}  & \colhead{6.2$\mu$m}  & \colhead{6.8$\mu$m}  & \colhead{7-9$\mu$m}  & \colhead{11.2$\mu$m}  &   \colhead{5.3$\mu$m}  & \colhead{6.2$\mu$m} & \colhead{6.8$\mu$m} & \colhead{7-9$\mu$m} & \colhead{11.2$\mu$m }&   & & &\cr      &\multicolumn{7}{c}{($10^{-17}~\rm W~m^{-2}$) } &
\multicolumn{5}{c}{(0.01\micron)} &
&\colhead{(\%)} &\colhead{(\%)} 
&\colhead{(\%)} 
}
\startdata
D1   &3060   
&3680
&1.0 &11&7.1 &27 &15    &0.02 &0.4 &0.2 &0.4 &0.4  &0.7  &3 & 2& 0.07 \\
D2   &67      
&110 
&0.9 &5.6 & 13  & 7.2  &2.9 &1.3 &5.5 &15&2.9 &2.3 &1.9 &50 & 30 & 0.7 \\
D4   &250     
&400
&1.0  &11 &1.4  & 5.6  &4.7 &0.4 &3.8 &0.6 &0.6&1.3 &2.4  & 20 & 10 &0.3 \\ 
D6   &67   
&110 
&0.8  &9.3 &13  &9.8&5.7 &0.9 &8.0 &17 & 3.4 &4.6 &1.6  & 70 &40 & 1 \\ Ellipse   &3500
&4740
&5.6 &46  &50  &63  &36 &0.2&1.3&1.4 &0.5&0.8 &1.3  & 7  & 8 & 0.2 \\ 
Chip-D1   &700
&1500
&8.8   &62   &85  & 81  &34  &1.2  &5.8  &10 &3.5  &2.6 &1.8 &50 &20  &0.5 \\ Chip    &3700 
&5100 
&12  &74  &113  & 86  &50 &0.3 &1.9 &3.0 & 0.6&1.0&1.5 &20 &8 & 0.2 \\ \enddata

\tablecomments{
$^a$ For the PAH analysis, D1 aperture is 0.8\arcsec;  D2, and D4  fluxes are for apertures of 0\farcs7 x 0\farcs6. Uncertainties in individual
feature fluxes are $\sim 20$-30\%. Fluxes for 
continuum-subtracted features are from modified Drude fits; fluxes from Gaussian
fits differ by $<10$\%. The `7\micron' and `14\micron' fluxes are
feature-free continuum fluxes in 1 \micron\ bands centered at 7
and 14.3\micron.
$^b$ Integrated line fluxes, in W m$^{-2}$.
 $^c$``PAH" is the sum of the fluxes of all fitted PAH features.
These are ratios of PAH 
to the continuum power listed in the first two columns. 
$^d$~Ratio of PAH feature fluxes to total estimated MIR flux, as from Table~\ref{tab:cdetails}, a 
rough estimate of the percentage of C in PAHs, see text. 
Absolute uncertainties for PAH to continuum ratios 
are estimated at 30-40\%.}
\label{tab:pahs}
\end{deluxetable*}

 {\subsubsection{PAHs and Dust; the Fraction of Carbon in PAHs}\label{subsubsec:pahs_fraccarbon}}
A measure of the relative populations of continuum-producing
very small dust grains or PAH clusters and PAHs can be quantified
by comparing to the mid-infrared fluxes and the infrared luminosity.

Following \citet{lebouteiller2007}, we define ``14\micron" as the continuum flux in the 1\micron\  wide featureless region centered at 14.3\micron. We  define a ``7\micron" flux similarly. The sum of all
the individual MIR PAH feature fluxes is designated `PAH'; it 
includes all features fit in Figure~\ref{fig:pah_plots}. The ratios
of PAH to the 7 and 14\mic\ fluxes are denoted ``PAH/7" and ``PAH/14" and are listed in Table~\ref{tab:pahs}.  PAH/7 and
 PAH/14 in D1, and the Ellipse and Chip regions dominated by D1, is much lower than in the other \hii\ regions,  $\sim $2-3\% in D1, versus 
 20-70\% in D2, D6, and Chip-D1. 
 Values for metal-rich galaxies are higher, with the lowest values occurring in high excitation environments  \citep{lebouteiller2011}; even the lowest
values in 30 Doradus, of similar metallicity to \gal, 
are several times higher than the value in D1. 

The percentage of carbon locked up in PAHs, $f_C$, can be estimated by comparing the summed spectral power, 
$L_{PAH} = \Sigma L_{feature}$ to the MIR luminosity, and using  
$f_{C} \sim L_{PAH}/L_{MIR}^{est}$ \citep{allamandola1989,tielens2008}.
Here we take the ratio of PAH power to $L_{IR}^{est}$, from the mid-infrared fluxes:
for D1, Ellipse, and Chip; this is a close estimate to the total infrared luminosity.
Values of $f_c$ for the regions in \gal\ are listed in Table~\ref{tab:pahs}. 
These values for $L_{PAH}$ are lower limits as they do not include the 3\micron\ feature. This feature could be strong, as  \citet{lai2025} found for the starburst in II~Zw~40, in which conditions may be 
 similar to D1. The fraction of power emitted in PAHs compared to the total
infrared power is  $\lesssim 1$\%  for all regions in \gal,
and lowest for D1, at 0.07\%, and low (0.2\%) for the
Ellipse and Chip regions dominated by D1. 
These values are much lower than values observed on larger scales in 
solar metallicity galaxies, which on average emit $\sim$10\% of their power
in PAHs \citep{smith2007}, or the 3-4\% calculated  for the Galaxy \citep{tielens2008, hrodmarsson2025}. 
The value for D1 is on the low end of the range observed for low metallicity dwarf galaxies, which average  
$\sim 0.5$\% \citep{hunt2010}.  D1 has $\sim$5-15 times 
less carbon locked up in the PAH features than do
the other \hii\ regions, D2, D4, and D6.   If 
one corrects for metallicity \citep{allamandola1989}, 
then the \fc\ values in 
Table~\ref{tab:pahs} should be multiplied by about three, 
which puts many of the regions in \gal\ closer to Galactic values, 
 but not D1.
\vspace{.2in}

\section{Discussion} \label{sec:discussion}

The \mrs\ instrument on JWST resolves the central starburst in \gal, isolates 
the giant cluster within cloud D1 from the nearby  less luminous \hii\ regions
D2 and D6, and enables the study of the effects of a luminous cluster on its surroundings.
Since D1, D2, D4, and D6 all share the same metallicity, it is
possible to distinguish to some degree the relative importance of metallicity versus local environment on the mid-infrared spectrum. We discuss whether metallicity or environment seem to dominate in individual aspects of the continuum spectra below. 

\subsection{Mid-infrared to Radio Continuum Ratio is Identical to the Galactic Value}\label{subsec:disc_mirradio}

Neither low metallicity nor intense radiation fields affect the
mid-infrared to radio continuum ratio in \gal. The excellent correspondence between hot MIR dust emission 
and radio emission in Galactic \hii\ regions is well established, and it holds in \gal\ as well.
At first glance it might seem surprising that the MIR/radio ratio
in the low metallicity \hii\ regions of \gal\ is the same as those in the 
Galaxy, since one might expect the gas-to-dust ratio, and thus
the dust emission, to scale with metallicity \citep{remy-ruyer2013}.
However, the bulk of the dust heating is due to  Lyman $\alpha$ absorption. So long as Lyman $\alpha$ is
trapped, this energy will be reradiated by hot dust within the \hii\ region. Although Lyman $\alpha$ does
not explain all of the dust emission, unless the IMF is pathologically
top-heavy  additional heating due to non H-ionizing photons should scale
with the Lyman continuum rate similarly to Galactic \hii\ regions.

\subsection{Mid-Infrared Spectral Index is Steeper in the Supernebula} \label{subsec:disc_mirspix}

To compare spectral slopes among the sources, we defined
spectral indices between 5, 15, and 20\micron. 
The spectral index values of  $\alpha_5^{15},~\alpha_5^{20}$
for supernebula D1 are 2-2.5, only slightly steeper than the less luminous \hii\ regions D2 and D6, 
which in turn are similar to values observed for the Orion Trapezium region.  
However the short wavelength spectral index, $\alpha_5^7$,
is significantly steeper in D1 than in the less luminous D2 and D6 \hii\ regions. 
According to the Galactic dust + PAH models of \citet{draineli2007}, 
the spectral indices of D1 as well as its spectral turnover at $\sim 20$\mic\ are consistent with $U\sim 10^5$-$10^6$, where $U$ is the standard interstellar radiation field.  The  
lower values of D2 and D6 combined with their longer
wavelength spectral turnovers are consistent with  only slightly 
lower excitation fields, $U\lesssim 10^4-10^5$. A strong 8\micron\ plateau in D1 could
explain the steep spectral index between 5 and 7\micron. Since they share the same metallicity, the
differences in mid-infrared spectral index between D1 and the other regions
are likely related to dust processing within D1. 

\subsection{PAH Features are Weak but are Seen Everywhere}\label{subsec:disc_pahsweak}
As expected for a low-metallicity environment \citep{engelbracht2008, madden2006, hunt2010, lai2020}, 
PAH features are weaker in all regions of the \gal\ central
starburst than in metal-rich environments. The EW of the PAH features
in the central starburst range from $\rm EW\sim 0.002-0.17$\micron. 
For comparison, the local solar-metallicity spiral, IC~342, has infrared luminosity and  star 
formation rate similar to \gal\ and EW of 0.5\micron\  \citep{brandl2006}.
 Even the moderate-metallicity (z$\sim 0.5 z_\odot$) \hii\ region NGC~5455 in
the outer part of M101 has equivalent widths of 
$\rm EW\sim0.8$-1\micron\ for the 6
and 11\micron\ features \citep{gordon2008a}. Taking the Chip-D1 region, which includes D2 and D6, 
 as  representative of the ambient star-forming ISM environment
 within \gal, 
 we find that the Chip-D1 region has EW$\sim 0.06, ~0.03$\micron\ for 
 the 6.2 and 11.2\micron\ bands, respectively. The very low
 EW across the central region of \gal\ 
 is consistent with the proposition that PAH
abundances drop sharply at metallicities below $\sim 0.2$-0.3 solar 
\citep{draineli2007,chastenet2019,shim2023}.

The highest PAH EWs are seen in regions D6 and Chip-D1, areas
relatively remote from D1. Region D4 was specifically defined to study the variation
of PAH features with distance from D1. The equivalent widths of its isolated PAH features are $\sim 0.04$-0.4\micron; the largest
EW is for the 6.2\micron\ feature, which is easily visible 
in D4. D4 contains the largest molecular cloud in the region and is located only 15 pc away from D1; 
the PAH emission may arise from cloud faces illuminated by D1 
but in a  less intense radiation field. Even with the
bright CO emission and largest molecular cloud, the PAH emission
is relatively weak in D4, the closest region to D1. 

\subsection{Solid State Features in D1 are Distinctive}\label{subsec:disc_D1distinct}

Within the central starburst of \gal, the compact, luminous D1 region 
is unusual even for a low metallicity, actively star-forming environment.
Intense or hard radiation fields weaken PAH emission 
\citep[e.g.,][]{lebouteiller2011, stock2014, onaka2015}.  
Region D1 has by far  the lowest PAH EWs  
among the \hii\ regions. The EWs of D1 are  
$\sim15$-20 times weaker at 6.2\micron\ and $\sim 2$-10 times weaker
at 11.2\micron\ than those of the less luminous D2 and D6 
\hii\ regions,  although the D1 feature fluxes of these bands
are as strong or stronger than those of D2 or D4. 
The comparatively low EWs in D1 result primarily 
from  the relative strength of the underlying MIR continuum at 7-9\micron, 
where the emission is dominated 
by very small dust particles (VSGs) or PAH clusters \citep{tielens2008}. 

The ratio of the  integrated intensities 
of the 6 and 11\micron\ features, $I_{6.2}/I_{11.2}$ 
(Table~\ref{tab:pahs}), can be an indicator of the importance
of charged relative to neutral PAHs \citep{tielens2008}, or
of the sizes of the PAHs, with smaller PAHs producing  relatively more 
6.2\micron\ emission. \citet{hunt2010} argue that low values are
evidence for a relative deficit of small PAHs at in low metallicity
galaxies. 
For all regions in \gal\ except D1, the ratio of 6.2\micron\ 
to 11.2\micron\ integrated intensity is 
$\sim 1.3$-2.4, within the
range of values observed for low metallicity galaxies  \citep{galliano2008, hunt2010}. 
In D1 the ratio is lower, 0.7.  A combination of the
hard radiation field and high intensities in 
D1 may be responsible for the 
low ratio, which could indicate large PAH sizes within the supernebula.

There are other distinctive aspects in the MRS 
spectra of D1. 
The 7.7\micron\ feature is
dominated by the relatively long wavelength component at 
7.86\micron\ in D1. 
The other \hii\ regions and ambient PAH emission have a stronger 7.6\micron\
component. The peak of the 7.7 complex is linked to PAH size, with larger
PAHS, $N_C\gtrsim 120$, producing emission at longer wavelengths 
\citep{vanDiedenhoven2004, bauschlicher2008, shannon2019}. 
As noted above, the ratio of the 6.2/11.2\micron\ integrated intensities
are also consistent with a larger particle size.

D1 also appears to have relatively less C locked up in smaller PAHs 
represented by these features as measured by its very low
equivalent widths, and low $f_C$, which is the
ratio of IR luminosity to PAH luminosity. The estimated
$f_C$ is $\sim 3$-10 times less in D1 than in the other
regions. This suggests that in D1, PAH
clusters or VSGs are relatively more abundant 
compared to the PAH population than in the surrounding regions.

A dramatic difference between D1 and the other regions is the strength 
of its 11.0\micron\ relative to 11.2\micron\ feature. The flux
of the 11.0\micron\ feature is $11\times 10^{-17}~\rm W\,m^{-2};$
the flux ratio $F_{11.0}/F_{11.2}=0.7$ in D1, an order of magnitude higher
than in the other regions. 
The 11.0\micron\ feature arises in cationic (charged)
PAHs, while 11.2\micron\ emission originates in neutral PAHs 
\citep{tielens2008, maragkoudakis2008}.
We have previously noted that the profile of the 11\micron\ feature in
D1 has a Class $B_{11.2}$ character, with an  extremely strong
11.0\micron\ wing that in D1 is nearly as strong as the 11.2\micron\ feature. Class B spectra  are associated with post-AGB, PNe, or Herbig Ae/Be stars 
\citep{vanDiedenhoven2004, tielens2008} rather than \hii\ regions;
D1 is  far too young to contain post-AGB stars or PNe,
although it could contain T Tauri, Herbig Ae/Be stars or young objects with similar dust processes. 

Features of PAH emission in D1 such as the very low EW, the low 
6.2/11.2\micron\ integrated intensity ratio, 
the longward shift of the
7.7\micron\ peak wavelength, and 
strength of the 11.0\micron\ feature taken together
point to a population of comparatively 
large, charged PAHs and a large proportion of continuum-emitting PAH clusters
or VSGs within the supernebula region, D1. It
is an open question whether this is due to the destruction of 
smaller or less stable PAHs, or PAH/grain growth in an environment
in which molecular clouds and hard, intense radiation fields are
in close proximity.

\subsection{The 6.8$\mu$m Feature: Possible Detection of Aliphatic 
Compounds within the Extended Starburst Region in NGC~5253} \label{subsec:disc_aliphatic}

The 6.8\micron\ feature that is so prominent in all spectra within
the starburst region of \gal\ is near a feature assigned to 
carbon skeletal stretching in aliphatic
structures \citep{esposito2024, fleming2026}, although this
feature, as many of these ``PAH" features, may originate
in mixed aliphatic-aromatic nanoparticles \citep{kwok2011, esposito2024}.  Astrophysical
sources dominated by the 6.8\micron\ feature  are classified as ``Class D"
\citep{matsuura2014}; they are carbon-rich and often post-AGB objects    \citep{sloan2014, jensen2022} and are perhaps linked
to dust production in these objects. 
The presence of this feature in \gal\ is a puzzle. 
The star clusters within D1, D2, or D6  cannot be old enough
to host significant numbers of AGB stars. While an extended population of 
evolved objects from previous starbursts could create this feature over the whole 100 pc field, it would not explain how it is also present in D1.

We suggest an alternate source of the 6.8\micron\ feature:  dust production
by Wolf-Rayet stars. \gal\ is known from ultra-violet and optical signatures to be a Wolf-Rayet galaxy \citep{crowther1999, schaerer1997}.  
JWST MIRI-MRS spectra of the WR binary star, WR 140,
by \citet{taniguchi2025} reveal a strong 6.8\micron\ 
feature toward both the star and its 
extended shells. The dust in WR 140 is produced in 
colliding winds within the binary system. 
We estimate that 100 such objects could conceivably produce the signal we see in \gal\, and D1 is sufficiently 
luminous to contain this number of WR stars.  The extremely compact environment
of D1, with closely spaced stars, may also favor soot production.

For this model to apply in \gal\ the macromolecules need to be dispersed by winds
from the central cluster, out to radii of 50 pc, 
with additional processing provided by the intense,
high excitation 
radiation fields \citep[e.g.,][]{chiar2013, andrews2015}.
Another possibility is that they are
produced in situ in the outer regions by heavily enriched,
super solar enriched \citep{tenoriotagle2005} 
winds from the cluster. 
Although the D1 cluster is  embedded and its CO
emission relatively undisturbed, it is
also known to be porous \citep{turner2017, consiglio2017, beck2025}. 
There is  evidence for outflows near D1  \citep{cohen2018, beck2020}.  
If the 6.8\micron\ feature is indeed related to WR stars in D1, it shows an intriguing
direction for the study of rapid metal enrichment in extreme, compact, low
metallicity starburst regions. 

\section{conclusions} \label{sec:conclusions}

The spatial and spectral resolution of the MIRI-MRS instrument
on JWST allow us to map the dusty starburst in \gal\  and isolate
its major mid-IR sources at parsec scales. A single pointing was observed, 
centered on the giant \hii\ region covers the surrounding
100 pc region. We describe the data analysis required by 
this extremely bright mid-infrared source and present
results on the continuum emission from the region surrounding
the giant \hii\ region.

\begin{itemize}
\item Four \mir\ continuum sources are detected in the central
100 pc starburst region of \gal. These coincide with 
submillimeter continuum and CO sources. 
Luminosities range from $10^7$ \lsun\ for fainter regions D2, D4, and D6, 
to $10^9$ \lsun\ for the dominant radio-IR \hii\ region, D1, the ``supernebula". 
The fainter \hii\ regions are comparable in luminosity to the brightest Galactic \hii\ regions. 

\item The spectra are roughly power-law in nature; there is a flat spectral component 
in most of the \hii\ regions  for $\lambda\lesssim 10$\mic, 
but a relatively steep rise from 7-10\micron\ for D1, suggesting  a 
 higher contribution of VSGs or plateau emission from PAH clusters.
The spectrum of D1 turns over beginning at $\lambda\sim20$\mic, similar to Galactic compact \hii\ regions. 

\item The mid-IR to radio continuum flux ratio for  supernebula D1
is the same as observed in the Galaxy: 
$S_{20\mu}/S_{2.8\rm mm}\sim 140.$ 
The primary fate of UV photons within the central 100 pc 
region of \gal\ is to emerge as 
\mir\ emission, which represents approximately half of the luminosity inferred
by free-free emission within the region. This does not include an additional 25\%
escape fraction suggested by more extended radio 
emission \citep{rodriguez-rico2007}. Direct 
absorption of UV photons by dust within the supernebula appears to be negligible, 
in spite of the high visual extinction. 

\item  The 9.7\mic\ silicate absorption feature is seen in aperture \gal-D1. It has an
optical depth of $\tau_{9.8}\sim 0.31\pm0.05$ as defined by the Spoon method.
This can be compared to
the observed visual extinction of $A_V \sim 35$. 
The 9.7\micron\ feature appears to be in emission in \gal-D2, as
it is in the Trapezium region of Orion \citep{cesarsky2000} as
well as the N66 region in the SMC \citep{whelan2013}. 

\item  PAH bands at 5.3, 6.2, 6.8, 7.7, 8.6, 11.2, and 12.7\micron\ are present
in all regions, but weak in this low metallicity dwarf galaxy. These
features account for $\lesssim 1$\% of the carbon atoms across
the region, and in D1, $\lesssim 0.1$\%.  
Among the four \hii\ regions, PAH fluxes are similar, but 
equivalent widths of the individual features differ by factors 
up to $\sim 50$. The largest EW are in region D6, relatively
remote from D1. 
The very low PAH equivalent widths of supernebula region D1 reflect
its relatively strong $\lambda<10$\micron\ continuum and/or plateau emission. 

\item  D1 has an unexpectedly high 11.0\micron/11.2\micron\ ratio; its
line profile is atypical for an \hii\ region and closer 
in its characteristics to  Class $B$ PAH emitters, which are often  evolved stars. 
A very strong 11.0\micron\ feature suggests 
an unusually large contribution from ionized PAHs. High 11.0/6.2 
ratios suggest that the source of this feature could be 
endo-PANHs \citep{ricca2021} in a galaxy with a well known nitrogen excess. 

\item The relatively low value of the 6.2/11.2 flux ratio in D1 could signify either a relatively low fraction of charged PAHs or larger PAH sizes compared to Galactic \hii\ regions. However, 
taken together with the high 11.0/11.2 flux ratio and the long peak wavelength of the 7.8\micron\ feature in D1, this points to large PAHs
 within the supernebula.

\item A strong 6.8\micron\ PAH feature likely to be from aliphatic 
compounds is present across the 100 pc starburst region.  
This feature is not typically 
observed in \hii\ regions, but is usually associated with post-AGB stars.
In \gal\ we suggest that this feature, as well as other
Class B features of the D1 spectrum, could be
caused by dispersal of the combined 
output of a population of sooty Wolf-Rayet stars.

\end{itemize}

\section*{Data Availability}

The JWST data presented in this article were obtained from the Mikulski Archive for Space Telescopes (MAST) at the Space Telescope Science Institute. 
The specific observations analyzed can be accessed via \dataset[doi: 10.17909/9mwc-xk11]{https://doi.org/10.17909/9mwc-xk11}.
The seven reduced the JWST MIRI/MRS spectra plotted in Figure \ref{fig:cspectra} are available as the data behind the Figure.

\begin{acknowledgments}
We are grateful for the excellent assistance we received in our many tickets
from the JWST help desk, including from David Law, Howard Bushouse, Boris Trahin, Kirsten Larson. We  thank the CARTA help desk for assistance, and
 Eric Greisen for the continuing existence of AIPS. We thank editor and
 the referee for a thoughtful and insightful review process
 that greatly improved the manuscript. We acknowledge with sadness the many contributions
of the late Sergiy Silich to our understanding of super star cluster
formation and his  
helpful comments on early versions of this manuscript, as well as many 
years of friendship and collaboration. His contributions to this field will be missed.
\end{acknowledgments}

\vspace{5mm}
\facilities{JWST(MIRI-MRS), IRSA, NED, ALMA}

\software{astropy \citep{astropy:2013,
astropy:2018,astropy:2022}, R \citep{Rcitation}, 
          Cloudy \citep{ferland2017}, 
          jdaviz \citep{jdadf_developers_2025_17211099},
          CARTA \citep{CARTA},
          AIPS \citep{greisen1990}
          }

\bibliography{N5253_JWST.bib}{}
\bibliographystyle{aasjournalv7}

\appendix

\section{Silicate Feature Fits} \label{sec:appendix_silicate}

Following the treatment of \citet{gordon2021}, the silicate feature properties were characterized using modified Drude functions. This function, 
similar to a Lorentzian, can be written as
$D_m =1/(1+z)^2$, where $z=[(\lambda/\lambda_0 - \lambda_0/\lambda)/
(\gamma/\lambda_0)]^2$. The width $\gamma = 2\gamma_0/(1+e^{(a(\lambda-\lambda_0)})$ is allowed to vary with wavelength, which permits the fitting of asymmetric profiles 
 through an asymmetry parameter, $a$ \citep{stancik2008}.
 
The underlying continuum for the 9.7\micron\ feature was extrapolated,
following the Spoon method \citep{spoon2007} by assuming 
power law and also 2nd order polynomials, with fits to the  5-6\micron\ and 14.5-15.2\micron\ regions of the loess-smoothed continuum on either side of the silicate feature. The power law fit had  residual standard error (RSE) 
$\sim 4$\% for these regions, the 2nd order polynomial 3\%. 
A fourth order polynomial gives a fit to the overall spectrum for 
the wavelength ranges 5-6\micron\ and 14-27\micron\ with a 
$\mathrm{RSE} \sim 6$\%. It does not fit the 5-15 \micron\ region as well.
  
The optical depth, \tausil, was determined by
taking the negative of the natural log of the ratio of the 9.8\micron\ flux interpolated
from the loess-smoothed data to the flux predicted by the assumed underlying 
continuum model. For the  2nd order polynomial continuum model, the value of  \tausil\  in region D1 is 0.23; the power law
model gives \tausil$\sim$0.31.   
Profiles of the silicate features were modeled in R by 
fitting the Drude function for the 
two features at 9.7\micron\ and 11.2\micron\ using  nonlinear least squares. The fit was done  for the spectral region between 7.8 and 12.5\micron.   RSE  
are $<7$\% for the fit region, dominated by the feature edges, where PAH contributions become important. The continuum models are shown
in Figure~\ref{fig:silicate}, along with the peak Drude wavelengths
and optical
depths at the Drude peak and 9.8\micron\  for the quadratic 
continuum model.

 \section{PAH Feature Fits}

  The first step in obtaining the PAH features was to determine
 a smooth PAH-free continuum, by which we use ``smooth" to
 mean features larger than 
 $\Delta\lambda\sim 0.5$\micron\ in extent.  The 9.7\micron\
 feature, which is much broader than the PAH features, was
 fit as part of the spline. 
 The spline fit was performed on the loess-smoothed line-free continuum following \citet{peeters2002} and \citet{chown2023}. We chose
 knots by hand, modifying the knots  given in analysis of \citet{chown2023} for the Orion PDR; 
 automatic knot-finding routines did not perform as well as manual weights,
 and did not reduce the number of knots required. An additional constraint
 was imposed that the spline must be less than the loess-smoothed continuum
 (positivity constraint). 
 The 41 knots used were [5.011, 5.042, 5.073, 5.15, 5.5, 5.85, 6.92, 7.25, 7.4, 8.3, 8.76, 9.15, 9.52, 9.6, 9.7, 9.8, 10.191, 10.344, 10.63, 10.7, 11.75, 11.95, 12.5, 13.5, 13.9, 14.45, 14.913, 15.178, 15.461, 16.0, 17.0, 18.5, 19.349, 19.632, 20, 21.0, 22, 23.7, 25.0, 26.0, 27] \micron. 
 The spline was excluded from the following wavelength regions with PAH features:
[(5.23, 5.32), (5.65, 5.82), (6.05, 6.91), (6.99, 7.14), (7.5,7.68), (7.73,8.0), (8.5, 8.75), (9.16, 9.495), (10.45, 10.6), 
(10.8, 11.57), (12.0, 12.4), (12.68, 13.0)]. 
In D1 and D4 the exclusion
region was modified to (6.05, 6.45), (6.55, 6.91) to allow the 6.5\micron\ 
continuum to be fit.

\begin{figure}
    \centering
    \includegraphics[width=\textwidth]{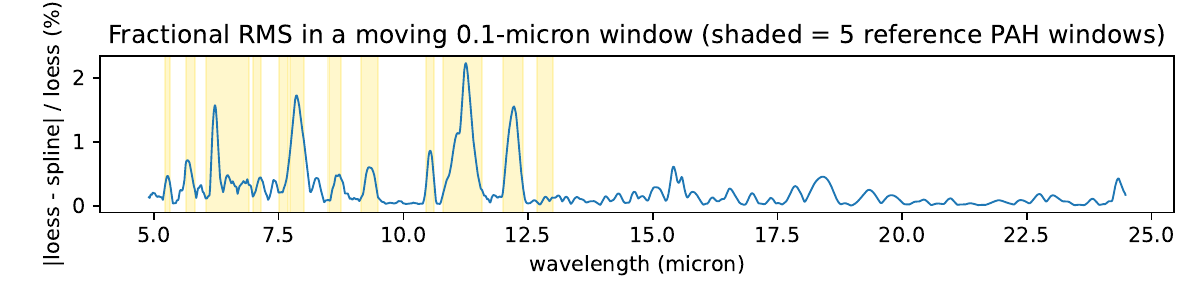}
    \caption{RMS percentage difference between loess-smoothed data and
    spline fit calculated for moving 0.1\micron\ window for the D1 spectrum. 
    PAH features are visible even in D1, in which they are very 
    weak. The peak is 2.2\% at 11.25\micron; the median value outside the PAH windows is 0.12\%.}
    \label{fig:movingrms}
\end{figure}

\begin{figure}
    \centering
    \includegraphics[width=5in]{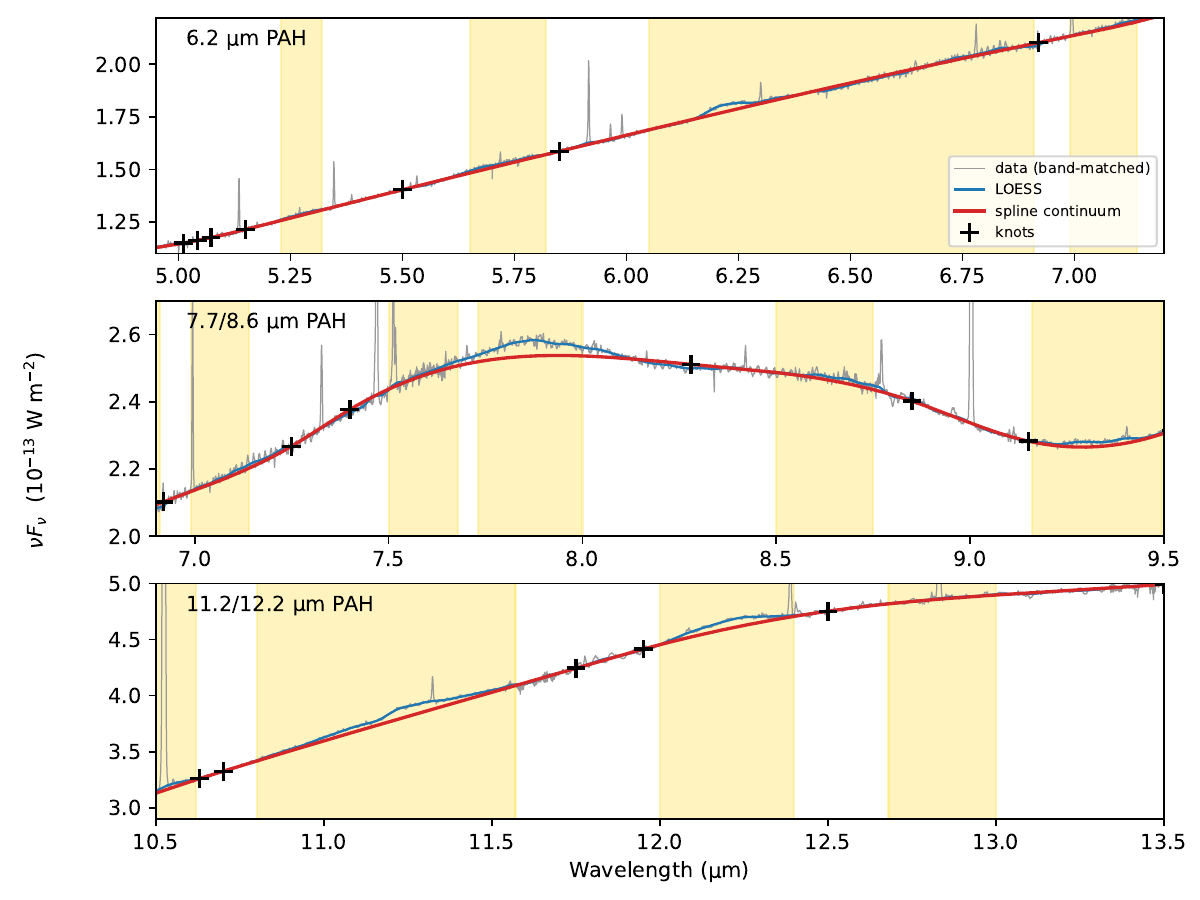}
    \caption{Spline fits for PAH spectral regions for the D1 spectrum. Its 
    very weak PAH features can be seen. 
    Yellow shaded wavelength regions were excluded from the spline fit, which 
    is depicted in the red line; crosses mark the knots.}
    \label{fig:splinefits}
\end{figure}

Once the spline was established,  features were obtained 
by subtracting the spline from the loess-smoothed line-free data.
The fractional 
rms of the loess-smoothed data minus the spline, represented as a moving percentage, is shown in Figure~\ref{fig:movingrms} for region D1.
Smoothed features can easily be seen in this figure. 
The median rms outside
the PAH windows ranged from 0.12\% (D1) to 1.1\% (D6). 
Spline fits with and without positivity constraints for the features were performed; the knot choices and PAH exclusions were 
iterated to minimize the rms of the difference spectrum
outside the PAH windows.
Positivity-constrained splines were adopted for the final splines.
The spline fits for the regions near the PAH features in the D1 region are shown in Figure~\ref{fig:splinefits}. 

Profiles were fit simultaneously to features
in three complexes from 6.0-9.2\micron,
10.9-11.6\micron, and 12.0-12.92\micron. 
Both Drude and Gaussian profiles were fit, 
using nonlinear least squares.  Drude profiles 
are appropriate for these macromolecules
\citep{leger1989, boulanger1998, draineli2007, smith2007}. 
The 6.2\micron\ and 11.2\micron\ feature profiles have a strongly asymmetric
 character that is better fit by modified Drude functions,
 which provide for asymmetry; we
 have employed the functional form presented in 
\citet{gordon2021}.  Residual standard errors (RSE) for
the Drude are $\sim 20$\% lower than the Gaussian fit RSE for the 6.2\micron\ and 11.2\micron\ 
features, confirming the visual impression (Fig.~\ref{fig:pah_plots}) that modified Drude functions
are  a better fit to the profile peaks. However, the leeward wings of these highly
asymmetric features are too broad. We briefly discuss individual complexes below.

{\it 5.3 \micron--- } Fluxes for the 5.3\micron\ feature were obtained
directly from the data using jdaviz-Specviz, for the 
spectral range corresponding to 5.22-5.32\micron, adopting a linear baseline.
This is only one of several features at 5-6\micron\ 
complicated in \gal\ by the presence of 
many faint nebular HI lines of the 
H7 series \citepalias{beck2025}.  
 
{\it 6--7 \micron--- } 
 Drude profiles were fit to 
features at 6.0, 6.2 and 6.8\micron. 
 Values for the asymmetry parameter, $a$, for
the strongly asymmetric 6.2\micron\ feature are $a= -24$ for D1,
and $a=-18$ to -28 for Ellipse, Chip-D1, and Chip. 
The 6.8\micron\ feature has the opposite asymmetry, $a=+30$ to +50.  
 RSE over the fit regions were 0.2\% for the higher flux regions
 such as D1, and $\sim 1.5$\% for D2 and D6. Fluxes measured over the fit
regions determined from the Drude profiles and  loess-smoothed data agree with the numerically 
integrated feature flux within the fit regions to $\lesssim 4$\%. 

{\it 7-9~\micron--- } This range is particularly difficult to fit 
since it overlaps with the blue edge of 
the 9.7\micron\ silicate feature and may include 
plateau emission from PAH clusters; 
the 8.6\micron\ feature lies
within the silicate feature. We did not attempt to 
fit the 9.7\micron\ feature simultaneously, 
but included it in the spline since it is much broader
than the PAH features. 
The spline fit in the  7.5-9\micron\ region is 
critical to both the absolute and relative 
intensities of the 7.7 and 8.6\micron\ features,
which depend sensitively on knot and exclusion choice.
We estimate that their ratios are uncertain to a factor of two,
which is why the 7-9\micron\ features 
are reported as one flux in Table 2. 
 Unlike the 6.2 and 11.2\micron\ features, these features are better fit 
 by Gaussians than Drudes, with RSE $<0.5$\%, but the difference
 is slight ($<$5\%). Feature widths are
 $\gamma_0=0.2$-0.22\micron, and 0.24\micron\ for D1. 
 Fluxes determined from Drude and Gaussian fits differ by less than 4\% from
 the integrated feature flux for the stronger regions and $\lesssim 7$\% for Chip-D1. Peak wavelengths of 
 the features, and the relative intensities of nearby 
 features, such as 7.6 and 7.8\micron, are robust to the 
 choice of spline fits. Fluxes have not been corrected for absorption by the  9.7\mic\ silicate
 absorption, which reduces the fluxes at 8.6\micron\ by $\sim$10-15\% in D1 and Chip.  

 {\it 11 \micron--- } 
Two features were fit in the 10.9-11.6\micron\ region,
at 11.0 and 11.2\micron. 
Like the 6.2\micron\ feature, the 11.2\micron\ profile 
 is strongly asymmetric and better fit with a modified Drude function. 
Asymmetry parameters were similar for all regions, 
$a=-26$ to $-37,$ with $a=-26$ for D1.  The width
parameter, $\gamma_0$, ranges from 0.11 to 0.17 in D1. RSE
are $<0.3$\%. Fluxes determined from Gaussian and Drude
fits differ by less than 4\% from the integrated feature
flux over the region.  RSE range from 0.2\% for D1 and
brighter regions, and $\sim 1.5$\% for D2, D6.

{\it 12 \micron--- }
Two features were fit within the  12.0-12.9\micron\ region. The first is
a strong feature at 12.2\micron. As discussed in the
text, the flux and width of this feature are consistent with that
expected from the 6.11\micron\ light leak 
\citep{wells2015, argyriou2023a}. The second
feature has fitted central wavelengths of 12.8\micron, which
coincides with the [Ne II] nebular line. Since
the nebular line is so narrow, it should
be weighted out of the fit, but there may be contributions
from faint line wings.
The fact that this PAH feature shows up more strongly in
Chip-D1 and D4 regions than in D1, unlike the nebular emission,
suggests that this is indeed the PAH feature and not
the nebular line.

\end{document}